\documentclass[preprint,floats,aps,12pt,superscriptaddress,nofootinbib,floatfix]{revtex4}

\usepackage[utf8]{inputenc}
\usepackage[T1]{fontenc}
\usepackage{lmodern}
\usepackage{amssymb,amsmath}
\usepackage{csquotes}

\usepackage[usenames,dvipsnames]{color}
\usepackage[normalem]{ulem}

\usepackage{hyperref}
\usepackage[separate-uncertainty=true,multi-part-units=single]{siunitx}
\usepackage{graphicx}
\usepackage[version=4]{mhchem}
\usepackage{textcomp}
\usepackage{placeins} \usepackage{natbib}

\usepackage[utf8]{inputenc}
\usepackage[T1]{fontenc}
\usepackage[english]{babel}
\usepackage{soul}

\UseRawInputEncoding

\begin{document}

\title{Non-equilibrium configurations of swelling polymer brush layers induced by spreading drops of weakly volatile oil}

\author{\"Ozlem Kap}
\affiliation{Physics of Complex Fluids Group and MESA+ Institute, Faculty of Science and Technology, University of Twente, PO Box 217, 7500 AE Enschede, the Netherlands}

\author{Simon Hartmann}
\affiliation{Institut f\"ur Theoretische Physik, Westf\"alische Wilhelms-Universit\"at M\"unster, Wilhelm-Klemm-Str.\ 9, 48149 M\"unster, Germany}
\affiliation{Center for Nonlinear Science (CeNoS), Westf{\"a}lische Wilhelms-Universit\"at M\"unster, Corrensstr.\ 2, 48149 M\"unster, Germany}

\author{Harmen Hoek}
\affiliation{Physics of Complex Fluids Group and MESA+ Institute, Faculty of Science and Technology, University of Twente, PO Box 217, 7500 AE Enschede, the Netherlands}

\author{Sissi de Beer}
\affiliation{Sustainable Polymer Chemistry Group, Department of Molecules \& Materials MESA+ Institute for Nanotechnology, University of Twente,PO Box 217, 7500 AE Enschede, The Netherlands}

\author{Igor Siretanu}
\affiliation{Physics of Complex Fluids Group and MESA+ Institute, Faculty of Science and Technology, University of Twente, PO Box 217, 7500 AE Enschede, the Netherlands}

\author{Uwe Thiele}
\affiliation{Institut f\"ur Theoretische Physik, Westf\"alische Wilhelms-Universit\"at M\"unster, Wilhelm-Klemm-Str.\ 9, 48149 M\"unster, Germany}
\affiliation{Center for Nonlinear Science (CeNoS), Westf{\"a}lische Wilhelms-Universit\"at M\"unster, Corrensstr.\ 2, 48149 M\"unster, Germany}

\author{Frieder Mugele}
\email{f.mugele@utwente.nl}
\affiliation{Physics of Complex Fluids Group and MESA+ Institute, Faculty of Science and Technology, University of Twente, PO Box 217, 7500 AE Enschede, the Netherlands}

\begin{abstract}

Polymer brush layers are responsive materials that swell in contact with good solvents and their vapors. We deposit drops of an almost completely wetting volatile oil onto an oleophilic polymer brush layer and follow the response of the system upon simultaneous exposure to both liquid and vapor. Interferometric imaging shows that a  halo of partly swollen polymer brush layer forms ahead of the moving contact line. The swelling dynamics of this halo is controlled by a subtle balance of direct imbibition from the drop into the brush layer and vapor phase transport and can lead to very long-lived transient swelling profiles as well as non-equilibrium configurations involving thickness gradients in a stationary state. A gradient dynamics model based on a free energy functional with three coupled fields is developed and numerically solved. It describes experimental observations and reveals how local evaporation and condensation conspire to stabilize the inhomogeneous non-equilibrium stationary swelling profiles. A quantitative comparison of experiments and calculations provides access to the solvent diffusion coefficient within the brush layer. Overall, the results highlight the -- presumably generally applicable -- crucial role of vapor phase transport in dynamic wetting phenomena involving volatile liquids on swelling functional surfaces. 
\end{abstract}

\maketitle

\section{Introduction}
Polymer brush layers consist of densely spaced polymer chains that are covalently attached at one end to a solid substrate. In dry state and in poor solvents, they form dense collapsed polymer layers on the substrate. Upon exposure to a good solvent, they swell. The degree of swelling is controlled by the balance of the osmotic pressure of the solvent and the elastic stretching of the polymer chains \cite{de1980conformations,milner1991polymer} and varies under the influence of many external stimuli such as temperature, pH value, solvent composition, electric fields, and illumination. This responsiveness can result in strong variations of many physical properties, including adhesion and fouling, friction and lubrication, mass transport and release with a wide variety of possible applications, as described in various review articles including Refs.~\cite{stuart2010emerging,yan2020brush, giussi2019practical, schubotz2021memory}. 
While most applications involve polymer brushes completely immersed in a solvent, recent years have seen an increasing interest in the wetting of polymer brushes and other soft materials, i.e., situations where responsive soft substrates are simultaneously exposed to the liquid solvent and to an ambient gas that is more or less saturated by solvent vapor~\cite{schubotz2021memory, butt2018adaptive, andreotti2020statics, ritsema2022fundamentals}. In particular, in dynamic situations where a drop of solvent is initially deposited onto a dry brush layer in a dry ambient atmosphere, this gives rise to a coupling between the spreading dynamics of the liquid, the evolution of the solvent vapor (for volatile liquid), and the swelling of the substrate with all the concurrent changes of its physical properties including the equilibrium contact angle. This specific responsiveness of polymer brush layers has been denoted as adaptive wetting \cite{butt2018adaptive}. Equilibrium properties of adaptive wetting systems, including also polyelectrolyte layers \cite{hanni2007water}, have been studied for quite some time and led to two persistent puzzles, namely Schroeder's paradox that adaptive wetting layers exposed to fully saturated solvent vapor are usually less swollen than upon immersion into bulk liquid and the fact that even good solvents often display partial wetting on brush layers, despite the -- by definition -- strong affinity between polymer and a good solvent \cite{cohen2006surfaces}. One additional challenge of adaptive wetting systems is that they often display multiple and very long relaxation times. This can make it difficult to judge whether `true equilibrium' is actually established in a given experimental situation. For instance, exposing polymer brushes to solvents of variable composition can lock in metastable molecular configurations that affect the wetting properties for months, as recently reported by Schubotz et al. \cite{schubotz2021memory} using a combination of contact angle measurements and sum frequency generation spectroscopy. 

The competition of different time scales becomes particularly evident in dynamic wetting situations when the intrinsic relaxation time scales interfere with the time scale of contact line motion that may be due to an externally imposed rate of change of the drop volume or arise from the intrinsic hydrodynamic spreading or evaporative retraction of the drop. Butt et al.~\cite{butt2018adaptive} recently pointed out the very general qualitative consequences of an intrinsically exponential contact angle relaxation process for the phenomenology of dynamic wetting experiments including, for instance, the appearance of contact angle hysteresis if the displacement rate of the contact line across the substrate is comparable to the relaxation time of the substrate (wettability) adaptation. To understand these processes for a specific system, it is essential to identify the actual relaxation processes involved in wettability adaptation and contact line motion. The spreading of drops on polymer brushes includes solvent transport by hydrodynamic drop spreading and solvent sorption by the brush layer. In the case of non-volatile solvents, the latter can only take place by sorption at the solid-liquid interface followed by imbibition of the solvent within the polymer brush layer. This process has been pictured either as a diffusive process of individual molecules~\cite{ThHa2020epjst} or as a hydrodynamic imbibition process like the imbibition of fluid into porous media \cite{shiomoto2018time, seker2008kinetics}. The latter gives rise to a liquid front that propagates with $x(t)\propto \sqrt(t)$ according to the classical Washburn law~\cite{washburn1921dynamics}. For volatile liquids, solvent evaporation, diffusion in the vapor phase, and subsequent condensation into the brush layer provide an additional pathway that can affect the coupled dynamics of drop spreading and swelling of the adaptive substrate. For inert solid substrates, the effect of evaporation and condensation on drop spreading has been studied extensively, see, e.g., Refs.~\cite{wayner1993spreading, shanahan2001condensation,CaGu2010sm,KoTS2014cocis,jambon2018spreading,WiDA2023arfm}. In this case, the competition between the divergence of both evaporation rate and viscous stress near the contact line leads to a complex scenario that results, for instance, in finite receding contact angles even for completely wetting liquids \cite{Morr2001jfm,ToTP2012jem}. For adaptive polymer brushes, the effect of vapor condensation might be even more important given the strong driving force arising for solvent sorption as initially dry brushes swell. At this stage, however, the role of evaporation and condensation on the dynamic wetting of adaptive substrates remains underexplored and poorly understood. This applies to the experimental perspective as well as to the one of modeling. For the latter, particular challenges arise from the need to incorporate multiple phases (liquid, vapor, dry polymer, swollen polymer) and their various transition and transport pathways. The resulting multi-scale aspects couple processes strongly localized near the three-phase contact line to the macroscopic dynamics of the bulk of the drop and the brush and vapor far away from the contact line. Further note that intricacies of contact line modeling are not limited to the wetting of polymer substrates but are related to fundamental questions in the physics of wetting~\cite{HuSc1971jcis,Duss1979arfm,Genn1985rmp,bonn2009wetting,SnAn2013arfm}. Similarly, the modeling of evaporation and condensation is related to fundamental questions of phase change dynamics, in particular, to the distinction of mass transfer across the interface limited by the actual phase change and by the diffusive transport of the vapour within the gas surrounding the drop \cite{PiBe1977jcis,SuBA2004jem,bonn2009wetting,WiDA2023arfm}, for a recent review see the introduction of \cite{HDJT2023jfm}. Of the wide range of approaches to the modeling of related dynamic phenomena, in particular, Molecular Dynamics simulations~\cite{mensink2019wetting,ritsema2020sorption,badr2022cloaking} and mesoscopic hydrodynamic models~\cite{ThHa2020epjst,GrHT2023arxiv} have been applied to the wetting of polymer brushes.

\begin{figure}
    \centering
    \includegraphics[width=\textwidth]{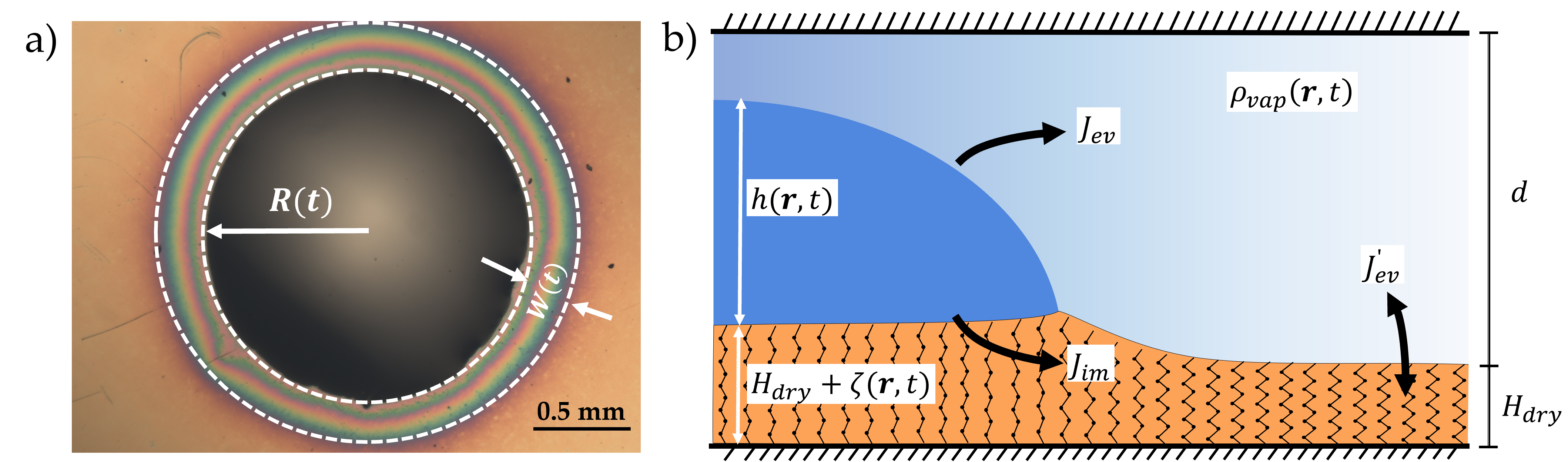}
    \caption{(a) Top view microscopy image of a hexadecane droplet after an hour of spreading on a PLMA brush layer; $R(t)$: drop radius; $W(t)$: width of the halo ahead of the macroscopic contact line. (b) Sketch of an evaporating drop on a polymer brush confined in a chamber of height $d$. The fields $h(r,t)$, $\zeta(r,t)$, and $\rho(r,t)$ represent the local liquid height, brush swelling and vapor saturation, respectively. Arrows indicate exchange fluxes between $h$, $\zeta$, and $\rho$. $H_\mathrm{dry}$ represents the dry thickness of a collapsed brush layer. Note that the relative sizes are illustrative and not to scale.}\label{fig:Fig1.png}
\end{figure}

In the present work, we study the spreading dynamics of drops of an oil, hexadecane (HD), with a low but finite vapor pressure and contact angle on a hydrophobic polymer brush layer of poly(lauryl methacrylate) (PLMA)~\cite{bielecki2013polymer,espinosa2013understanding} and the resulting inhomogeneous swelling dynamics of the adaptive substrate formed by the brush layer. Using video imaging and microscopic interferometry, we quantify the macroscopic spreading dynamics and demonstrate the emergence of a halo of partially swollen brush layer ahead of the moving contact line in the later stages of the spreading process (Figure~\ref{fig:Fig1.png}). This halo can reach extensions of several hundred micrometers on a time scale of several hours and can assume different long-living i.e., quasi-stationary, non-equilibrium configurations depending on the containment of the evaporating solvent vapor. A gradient dynamics model for the evolution of three independent fields is developed and numerically solved. It reproduces the temporal evolution of the halo and provides insights into the relative importance of competing transport mechanisms through the vapor and within the brush layer.

\section{Materials and Methods}\label{sec:exp_setup}

\subsection{Chemicals}
Silicon wafers (\SI{100.0 \pm 0.5}{mm} diameter and \SI{525\pm25}{\um} thickness, boron-doped with (100) orientation, \SI{510}{\ohm cm}, Okmetic) were cut into $2\times\SI{2}{cm^2}$ pieces for characterization and synthesize.
Lauryl methacrylate (LMA, 96\%, CAS 142-90-5), copper(II) chloride (\ce{CuCl2}, 97\%, CAS 7447-39-4), $\alpha$-bromoisobutyryl bromide (BiBB, 98\%, CAS 20769-85-1), N,N,N',N'',N''-pentamethyldiethylenetriamine (PMDETA, 99\%, CAS 3030-47-5), triethylamine (TEA, 99\%, CAS 121-44-8), (3-aminopropyl) triethoxysilane (APTES, 99\%, CAS 919-30-2), ascorbic acid (>99\%, CAS 50-81-7), sulfuric acid (\ce{H2SO4}, 98\%, CAS 7664-93-9), hydrogen peroxide (\ce{H2O2}, 30\%, CAS 7722-84-1) were purchased from Sigma-Aldrich, toluene (99.8\%, CAS 108-88-3) was purchased from Alfa Aesar, and n-hexadecane (99\%, CAS 544-76-3) was purchased from Acros Organics and used as received without purification. Ultrapure water (resistivity \SI{18.2}{M\ohm cm}) was obtained from a Millipore Synergy UV system.

\subsection{Polymer brush synthesis and characterization}
The oxidized Si wafers were functionalized with bottle brushes of poly(lauryl methacrylate) (PLMA), i.e., a polymer with a polymethacrylate backbone functionalized with fully saturated lauryl side chains that provide a hydrophobic character. Surface functionalization was conducted in a grafting-from approach employing the surface-initiated activators regenerated by electron transfer atom transfer radical polymerization (SI-ARGET-ATRP). This method requires little (typically ppm) metal catalyst and provides better oxygen tolerance compared to conventional ATRP methods~\cite{min2007use, kwak2011arget}. Three pre-functionalization steps (surface hydroxylation, silanization and initiator coupling) were performed following standard procedures as described in the literature~\cite{brio2020degrafting} before starting the actual polymer brush synthesis. 
The specific SI-ARGET-ATRP recipe was adapted from Ref.~\cite{dunderdale2015polymer} with minor adjustments to the reactant ratios. Ascorbic acid (AA) (\SI{40}{mg}, \SI{227}{\micro mol}) and ethanol (\SI{3.5}{mL}) were mixed in a glass vial (\SI{10}{mL}, \SI{2}{cm} diameter). CuCl\textsubscript{2} (\SI{28}{mg}, \SI{210}{\micro mol}) and PMDETA (\SI{100}{\micro L}, \SI{480}{\micro mol}) were mixed in ethanol (\SI{10}{mL}). A volume of \SI{0.5}{mL} Cu catalyst solution was added to the glass vial containing AA. Monomer (\SI{4}{mL}, \SI{13.65}{mmol}) was added to the vial, and the mixture was stirred. The initiator-modified substrate was inserted into the reaction solution, and the glass vial was sealed with a screw-top lid. Reaction solutions were not degassed, and glass vials contained $\sim$\SI{4}{cm^3} volume of ambient air. After 3 hours of reaction time, the substrates were rinsed with toluene, water and ethanol and dried with a nitrogen stream.

\subsection{Characterization methods}\label{sec:characterization}
The dry thickness of the polymer brushes $H_\mathrm{dry}$ was measured to range between 180 and 220 nm using a Spectroscopic Ellipsometer (SE) with Nanofilm-EP3 SE (ACCURION GmbH, G\"ottingen, Germany) at angles of incidence of \SI{60}{\degree}, \SI{65}{\degree} and \SI{70}{\degree} in a spectral range of \SI{400}{nm} to \SI{995}{nm}.  
Optical images of the spreading drops were recorded using an upright microscope (Nikon Eclipse, L150) with a color camera (Basler a2A5328 - 15ucBAS). The macroscopic spreading behavior was quantified by imaging under white light illumination. Quantitative information about local swelling profiles was obtained using interferometric imaging under monochromatic illumination with a narrow band green filter ($\lambda = \SI{532 \pm 10}{nm}$ Thorlabs, FL05532-1). More detailed information about the analysis steps are provided in the Supporting Information (SI) (Figure~(\ref{fig:SI_1})).

\subsection{Theoretical model}
\label{sec:model}
The theoretical description of the system is based on the framework of gradient dynamics as employed in the mesoscopic hydrodynamic modeling of complex wetting \cite{ThTL2013prl, ThAP2016prf, Thie2018csa}. In particular, we extend an earlier model by Thiele and Hartmann \cite{ThHa2020epjst} for a non-volatile liquid on a polymer brush. The system is described employing an underlying a free energy functional $F[h,\zeta,\rho_\mathrm{vap}]$ that depends on three independent fields, namely, the thickness of the oil layer \(h(\vec{r},t)\), the excess brush thickness due to the local degree of swelling \(\zeta(\vec{r},t) \), and the local vapor density \(\rho_\mathrm{vap}(\vec{r},t)\) (Figure~\ref{fig:Fig1.png}). Here, \(\vec{r}=(x,y)\) and \(t\) are the substrate coordinates and time, respectively. 
While the model is presented in the general form below, in all the numerical calculations, we only consider radially symmetric geometries.
Moreover, we assume that the extension of the experimental chamber in the vertical direction is small as compared to its horizontal dimensions such that the vapor quickly equilibrates in the vertical direction  and \(\rho_\mathrm{vap}\) can be considered to only depend on \(\vec{r}\) and \(t\). A detailed assessment of this approach can be found in Ref.~\cite{HDJT2023jfm}.

Then, the free energy \(F[h,\zeta, \rho_\mathrm{vap}]\) corresponds to
\begin{align}
    F = \int_\Omega \Bigg[
        \underbrace{\gamma_\mathrm{lg} {\sqrt{1+|\nabla (h+\zeta)|^2}}}_\text{liquid-gas\,interface\,energy}
        + \underbrace{\gamma_\mathrm{bl}(\zeta)\sqrt{1+|\nabla \zeta|^2}}_\text{liquid-brush\,interface\,energy}
        + \underbrace{f_\mathrm{wet}(h,\zeta)\sqrt{1+|\nabla \zeta|^2}}_\text{wetting potential}\nonumber\\
        + \underbrace{f_\mathrm{brush}(\zeta)}_\text{brush energy}
        + \underbrace{(h+\zeta) f_\mathrm{liq}(\rho_\mathrm{liq})}_\text{liquid bulk energy}
        + \underbrace{(d-h-\zeta) f_\mathrm{vap}(\rho_\mathrm{vap})}_\text{vapour energy}
        + \underbrace{(d-h-\zeta) f_\mathrm{air}(\rho_\mathrm{air})}_\text{air energy}
        \Bigg] \mathrm{d}^2x,\label{eq:free-energy-functional-main}
\end{align}
where \(f_\mathrm{liq}\), \(f_\mathrm{vap}\), and \(f_\mathrm{air}\) are bulk liquid, vapor, and air energies per volume, that are converted to energies per substrate area by multiplication with the effective liquid height $(h+\zeta)$ or local gap height $(d-h-\zeta)$. Furthermore, \(\gamma_\mathrm{lg} \) is the constant liquid-vapor interface energy, \(\gamma_\mathrm{bl}\) is the brush saturation-dependent liquid-brush interface energy, and \(f_\mathrm{wet}(h,\zeta)\) is the brush saturation-dependent wetting energy per unit area. Also, \(f_\mathrm{brush}(h,\zeta)\) is the Flory-Huggins-type energy of the partially swollen brush containing an elastic and an entropic contribution. (For the present system of alkyl-terminated bottle brushes wetted by a pure alkane, the Flory-Huggins $\chi$-parameter is chosen to be zero.) Adaptivity of the equilibrium wettability of the system arises from the dependence of  \(f_\mathrm{wet}\) on the local degree of swelling, i.e., on $\zeta$. Detailed expressions for each term are provided in the Appendix.

Variation of the free energy with respect to \(h\), \(\zeta\), and \(\rho\) yields the corresponding three chemical potentials. Taking the conservation of the number of molecules of the fluid across all phases into account, the time evolution of each field at any position can be written as the sum of a conserved flux driven by gradients of the corresponding chemical potential and non-conserved fluxes \(J_i\) arising from the transfer of particles between the different fields due to evaporation ($i=\mathrm{ev}$) and imbibition ($i=\mathrm{im}$).

Simplifying the expressions and replacing the local degree of swelling \(\zeta\) by the dimensionless swelling ratio $\alpha=(H_\mathrm{dry}+\zeta)/H_\mathrm{dry}=1+\zeta/H_\mathrm{dry}$ the resulting dynamic equations read
\begin{equation}
    \begin{aligned}
        \partial_t h           & = \nabla\cdot \left[ \frac{h^3\rho_\mathrm{liq}}{3\eta} \nabla \mu_\mathrm{liq} \right] - J_\mathrm{ev} - J_\mathrm{im} \\
        \partial_t \alpha      & = \nabla\cdot \left[ \frac{D_\mathrm{brush}}{k_BT} \, (\alpha-1) \, \nabla \mu_\mathrm{brush} \right] + \frac{1}{H_\mathrm{dry}}(J_\mathrm{im} - J_\mathrm{ev}') \\
        \partial_t [(d-h)\phi] & = \nabla\cdot \left[D_\mathrm{vap} (d-h) \nabla \phi \right] + \frac{\rho_\mathrm{liq} k_B T}{p_\mathrm{sat}}(J_\mathrm{ev} + J_\mathrm{ev}').
    \end{aligned}\label{eq:dynamic_eqs_final-main}
\end{equation}

Here, $J_\mathrm{im}$, $J_\mathrm{ev}$, and $J'_\mathrm{ev}$ are non-conserved fluxes that describe the transfer of oil between the three fields, namely, transfer by imbibition from the bulk liquid into the polymer layer, transfer by evaporation/condensation between bulk liquid and vapor phase, and transfer by evaporation/condensation between the partly saturated brush layer and the vapor phase. Note that from now on, we only consider radially symmetric geometries and employ $r$ as radial coordinate. A detailed description of the model, derivations of the relevant equations, and the values of all parameters are provided in the Appendix.

\section{Results and Discussion}\label{sec:results}

\subsection{Macroscopic spreading dynamics}
\begin{figure}[hbt!]
    \centering
    \includegraphics[width=\textwidth]{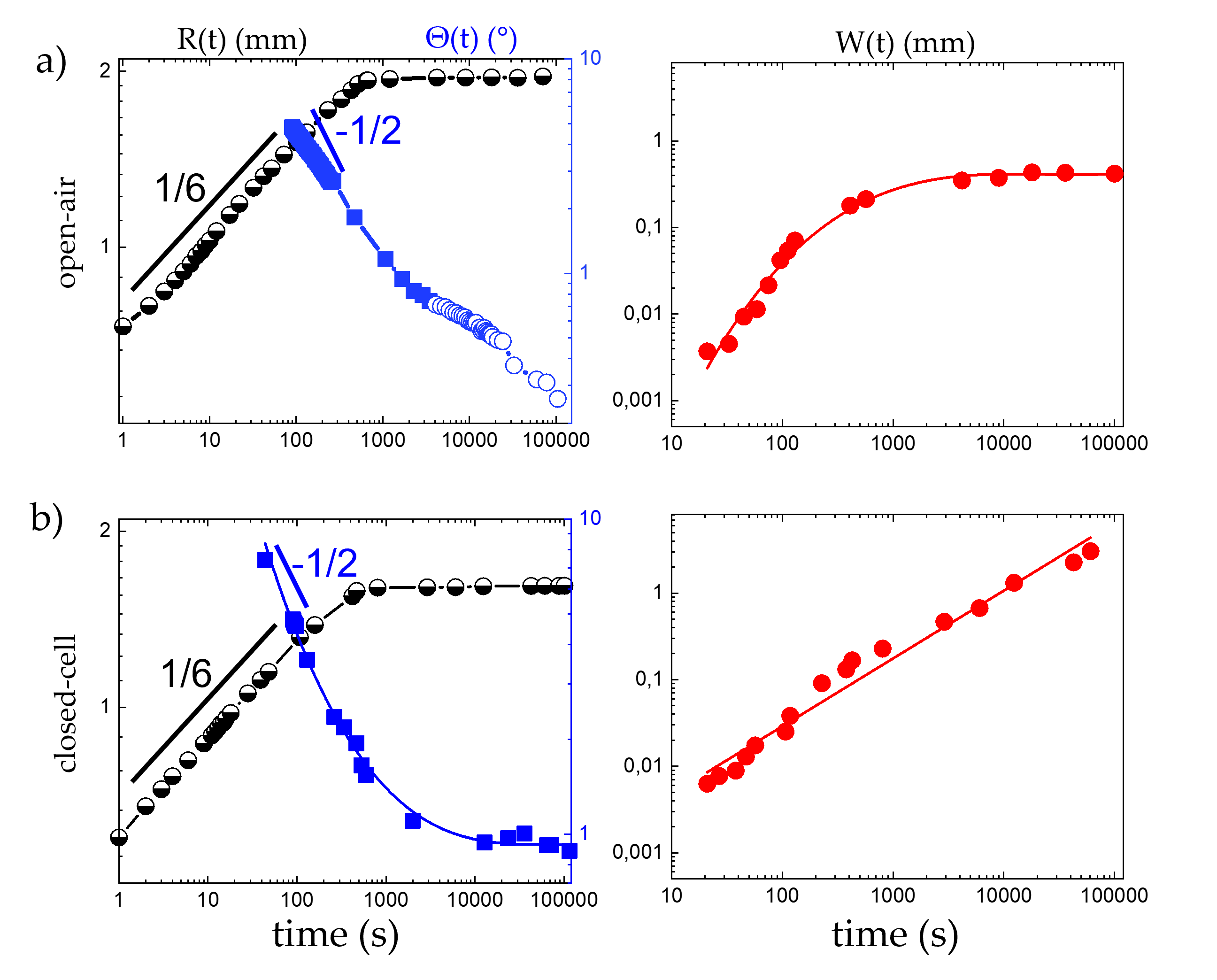}
    \caption{{Characterization of macroscopic drop evaporating and spreading on PLMA brush layer a) in the open configuration, and b) in the closed configuration. Left panels: drop radius $R(t)$ (black symbols) and contact angle $\theta(t)$ (blue symbols; open blue symbols in the top left panel are affected by slight contact angle hysteresis; see text for details). Right panels: halo width $W(t)$.}}
    \label{fig:Fig2.png}
\end{figure}

\begin{figure}[hbt!]
    \centering
    \includegraphics[width=\textwidth]{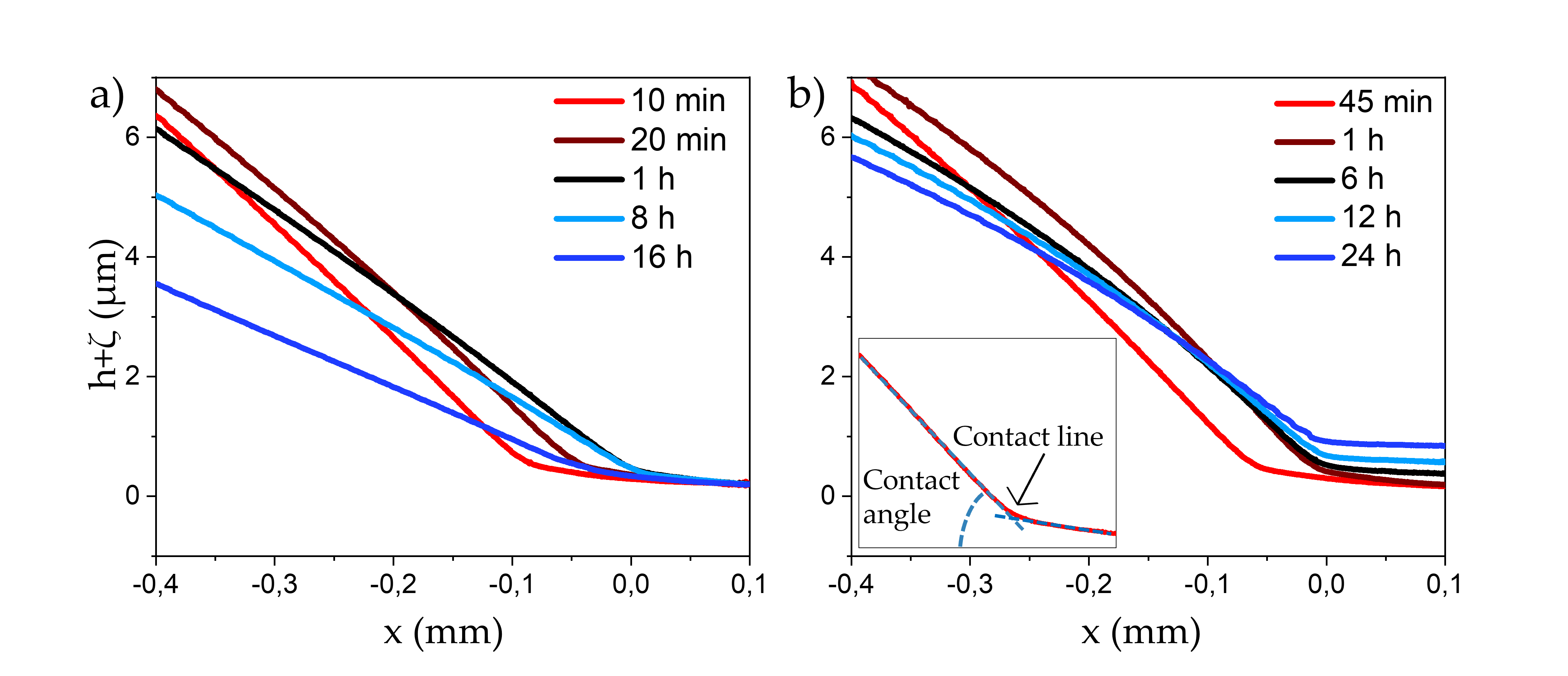}    \caption{Time-dependent droplet height profile in the contact line region obtained from  interferometry. (a) open cell and (b) closed cell configurations. The inset in (b) gives a contact line region and illustrates the extraction of the contact line position (by linear extrapolation) and contact angle.} 
    \label{fig:Fig3.png}
\end{figure}

Oil drops are deposited onto the polymer brush substrate to spread under two different conditions. In the \emph{open configuration}, the samples are mounted in a sample cell open to the ambient air. In the \emph{closed configuration}, we close the sample cell within seconds of depositing the drop by placing a microscope cover slip to contain any vapor of evaporated liquid. In both situations, top-view video images allow us to extract the drop radius $R$ as a function of time. For both configurations, $R$ initially increases algebraically with time as $t^\nu$ and an exponent of $\nu\approx1/6$ (Figure~\ref{fig:Fig2.png}). Contact angles extracted from droplet height profiles (Figure~\ref{fig:Fig3.png}) using interferometry images which are recorded with the same conditions, show that $\theta$ decreases algebraically with an exponent $\mu\approx-1/2$.  As expected, the values of $\nu$ and $\mu$ are consistent with the elementary geometric relation $r\propto\theta^{-3}$ for spherical caps of fixed volume for $\theta\ll1$ as valid at short times. After approximately 15--20\, min, the spreading process saturates, and the macroscopic drop shape approaches a nearly stationary state for both open and closed configurations. (For the open configuration, the contact angle keeps decreasing long after the radius has saturated (open blue symbols in the left panel of Figure~\ref{fig:Fig2.png}a). We attribute this continued decrease to a combination of gradual drop evaporation and a small contact angle hysteresis of approximately \SI{0.5}{\degree}. 

The numerical values of $\nu$ and $\mu$ deviate from the classical exponents $\nu_T=1/10$, and $\mu_T=-3/10$ given by Tanner's law that describes the spreading of non-volatile Newtonian liquids on solid substrates with a perfect no-slip boundary condition \cite{tanner1979spreading}. Qualitatively, this is not surprising. The interface between the swollen brush and the bulk drop is rather diffuse, and displays dilute, flexible polymer chains that are easily deformed by the strong viscous stresses close to the contact line. Both, the diffuseness and the possibility of local shear thinning or slip, will apparently lead to an effective hydrodynamic boundary condition that alleviates the stress divergence and thereby promotes faster spreading than in Tanner's law \cite{bonn2009wetting}. Moreover, local evaporation and condensation also affect fluid transport \cite{wayner1993spreading}.

At first glance, one might also be surprised that the two different forms of vapor containment lead to the same type of macroscopic spreading behavior regarding drop radius and contact angle. This arises from the fact that the brush layer is initially dry in both cases. A significant difference in the spreading behavior can only be expected once the system has time to experience the difference in the boundary conditions for the vapor. At the very least, molecules in the vapor must have had enough time to diffuse to the edge of the experimental cell. For a cell diameter of a few centimeters, this is the case after a characteristic diffusion time $T_\mathrm{diff}=L^2/D_\mathrm{vap}$, which amounts to about ten seconds for a vapor diffusion coefficient $D_\mathrm{vap}=\SI{e-5}{m^2/s}$ for hexadecane in air. 

To illustrate that the swelling state of the brush layer does indeed affect the spreading behavior, we performed spreading tests on brush layers that were pre-equilibrated in saturated HD vapor inside the closed chamber for up to three weeks. This leads to homogeneous pre-swelling of the brush layer by a factor of $\approx 2$ compared to the dry thickness. The chamber is then quickly opened to deposit an HD drop and immediately closed again. The subsequent spreading of the drop results in a slower algebraic drop spreading with an exponent of $\nu_\mathrm{sat}\approx1/8$  (Figure~\ref{fig:SI_2}). Pre-swelling thus clearly affects the macroscopic spreading dynamics in our system, similar to earlier reports for polyelectrolyte layers~\cite{hanni2007water}.

\subsection{Halo evolution}
\begin{figure}[hbt!]
    \centering
    \includegraphics[width=\textwidth]{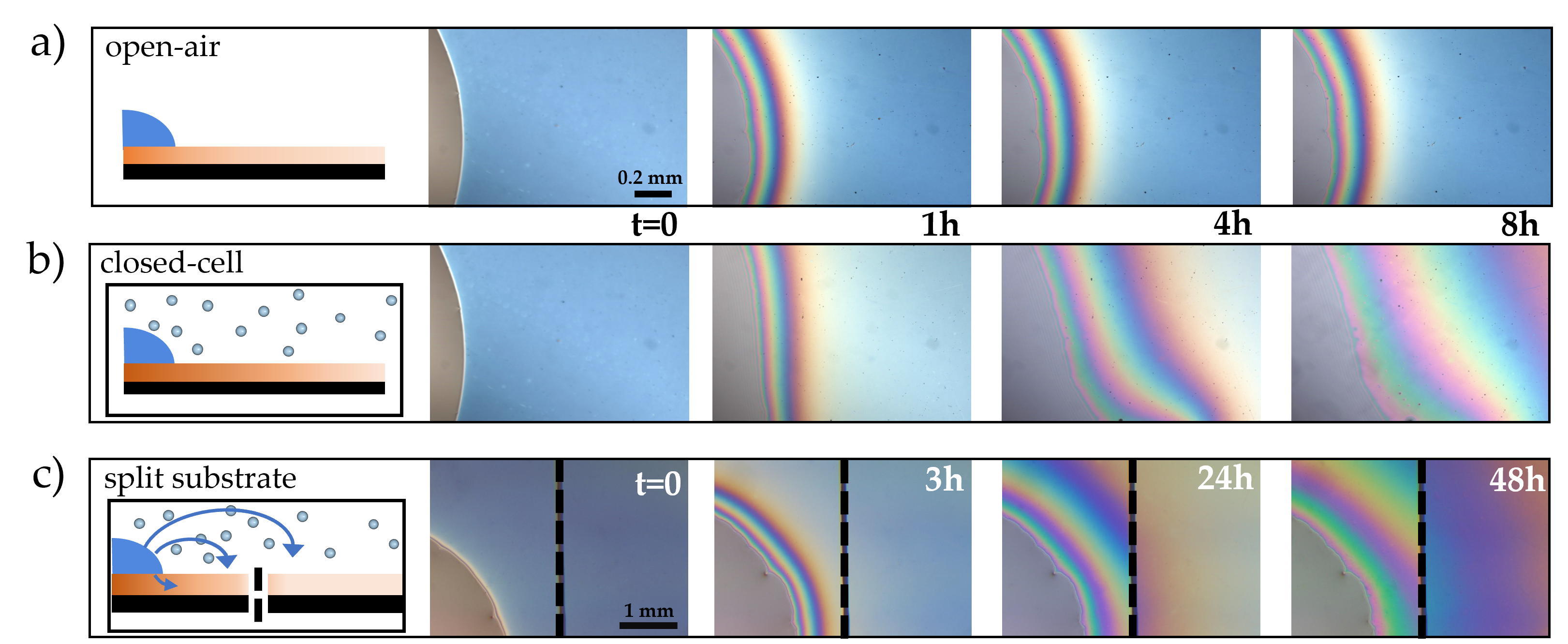}
    \caption{Illustrating sketches and optical images of a hexadecane droplet (grey) spreading on a PLMA brush (coloured): Relevance of vapor containment and substrate configuration for drop spreading and brush swelling. a) drop spreading in the open configuration with a finite halo width in the stationary state. b) In contrast, drop spreading in the closed configuration induces a continued expansion of the halo. c) Spreading as in b) but on the split substrate.}
    \label{fig:Fig4.png}
\end{figure}

    Of primary concern in the present work is, however, not the macroscopic spreading behavior of the drop but the effect of drop spreading on the swelling of the polymer brush layer. Immediately after deposition, the drop quickly spreads across the dry polymer brush layer (see the video in the Supplementary Information). After only a few tens of seconds, a colorful halo emerges, indicating that a zone of partly swollen polymer brush layer appears ahead of the moving contact line. While the initial development of the halo is independent of the vapor containment, its subsequent behavior at long times is very different: in the open configuration, the halo initially extends its width $W$ but then saturates after 15-20 min, right panel of Figure~\ref{fig:Fig2.png}a. In contrast, in the closed configuration, $W$ grows indefinitely, right panel of Figure~\ref{fig:Fig2.png}b. Then, at a very late stage, its outermost edge becomes somewhat 'wavy', rendering its exact width difficult to determine. The difference between the two configurations becomes very clear from magnified images of the contact line region. They are given in Figure~\ref{fig:Fig4.png}a \& b and very clearly show how the halo assumes a stationary state in the open configuration while it continues to widen in the closed one.
   
\begin{figure}[hbt!]
    \centering
    \includegraphics[width=\textwidth]{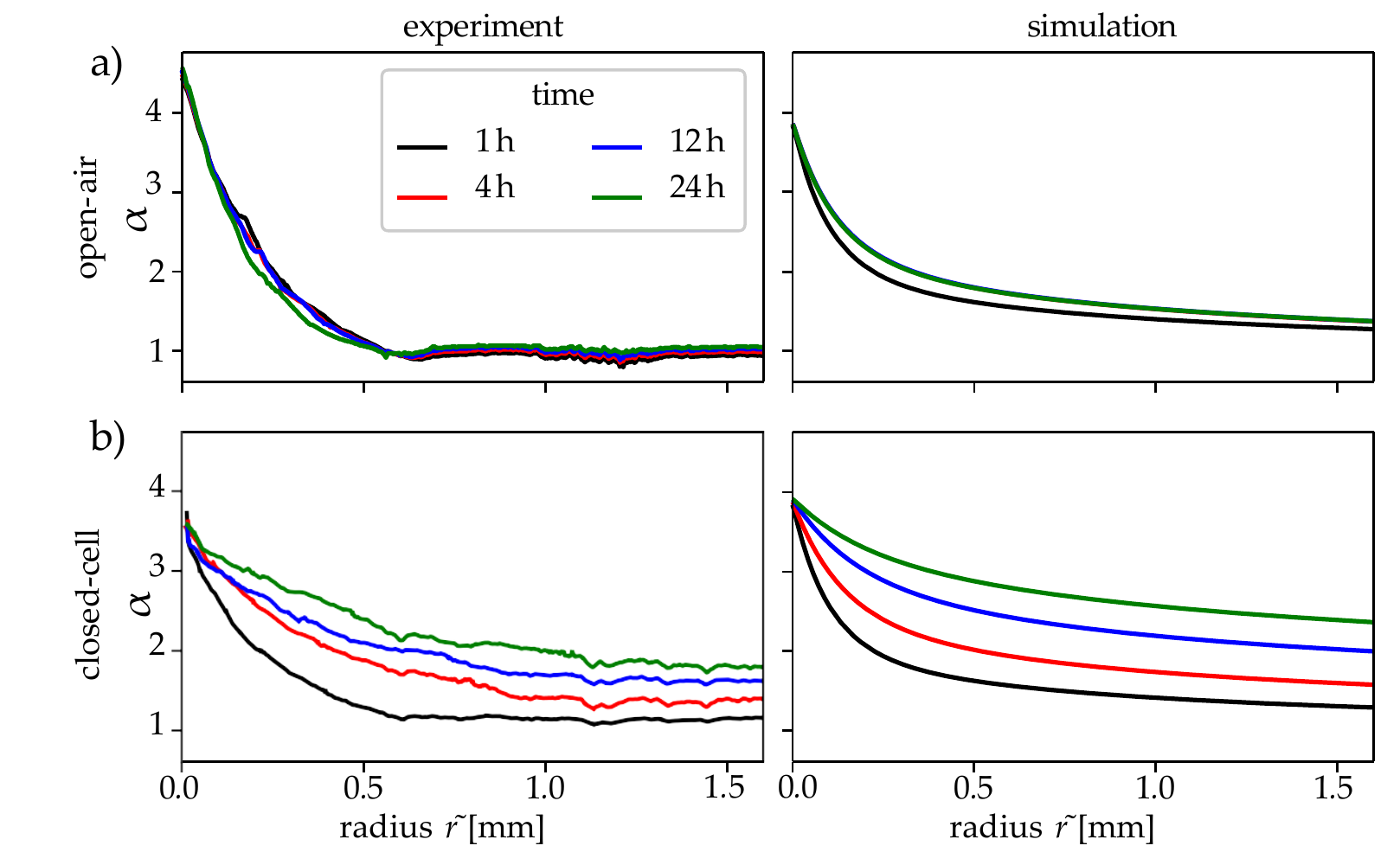}
    \caption{Brush swelling ratio profiles $\alpha(r,t)$ are given a) in the open configuration, and b) in the closed configuration at various times (black: 1\,h; red: 4\,h; blue: 12\,h; green 24\,h) as a function of the radial distance to the contact line $\tilde r=r-R$. Left: experimental data ($H_\mathrm{dry} = \SI{180}{nm}$). Right: numerical results. Note that in the open air case the curves converge after a short time.}
    \label{fig:profiles}
\end{figure}

The same behavior is seen in the brush swelling ratio profiles $\alpha(\tilde r,t)=h(\tilde r,t)/H_\mathrm{dry}$ (Figure~\ref{fig:profiles}) that we extract from the analysis of the monochromatic interferometry images. Note that, here, $\tilde r=r-R$ is the radial distance to the contact line. In the open configuration, these profiles converge onto a universal curve for $t\ge\SI{1}{h}$ with a maximal swelling ratio of nearly 5 close to the contact line at $\tilde r=0$.
Far away from the contact line, the film remains in its dry state with $\alpha=1$ at all times. In contrast, in the closed configuration, the profile does not converge but continues to evolve even on our maximal experimental time scale of 24\,h. While the maximum of swelling ratio close to the contact line remains nearly constant at a value of about 4 -- only slightly smaller than in the open configuration -- the brush layer continues to swell across the entire sample. Even far away from the contact line, after 24\,h, the swelling ratio reaches values up to 2. The comparison between the open and the closed configuration thus clearly proves that vapor phase transport is crucial for the spreading-induced swelling of the brush layer, despite the very low vapor pressure of HD. 

To explicitly demonstrate the simultaneous contributions of liquid imbibition and vapor phase transport, we perform additional experiments with a substrate purposefully broken into two pieces. Within the chamber, the two parts of the substrate are then placed next to each other, separated by a small gap as indicated by the black dashed lines in Figure~\ref{fig:Fig4.png}c. A drop is deposited onto the left piece, the cell is closed, and the spreading process is observed. As the drop spreads, as expected, a halo develops close to the contact line. After a few hours in the closed cell, the brush layer also starts to swell on the right piece. Yet, comparing the color variation far away from the contact line on the two separated pieces, it becomes clear that the brush layer on the left-hand piece swells more quickly than the one on the right-hand piece. From this observation, we conclude that the brush swells faster if it is simultaneously fed by both direct liquid imbibition and condensation from the vapor phase. In contrast, the right-hand piece still shows significant but slower swelling as it is only fed via oil condensation from the vapor phase. This experiment thus unambiguously demonstrates that in the present system, both transport mechanisms operate in parallel and that they are both of appreciable importance. It remains an intriguing observation, though, that the brush layer in the open configuration assumes a stationary state featuring a pronounced gradient in brush swelling ratio once the macroscopic spreading process has saturated. Such gradients in a stationary state are incompatible with thermodynamic equilibrium and can only exist in the presence of persistent gradients in chemical potential. Despite their longevity, the observed brush profiles must therefore reflect a stationary ongoing non-equilibrium process in the system.

\subsection{Modeling results} 

\begin{figure}[hbt!]
    \centering
    \includegraphics[width=\textwidth]{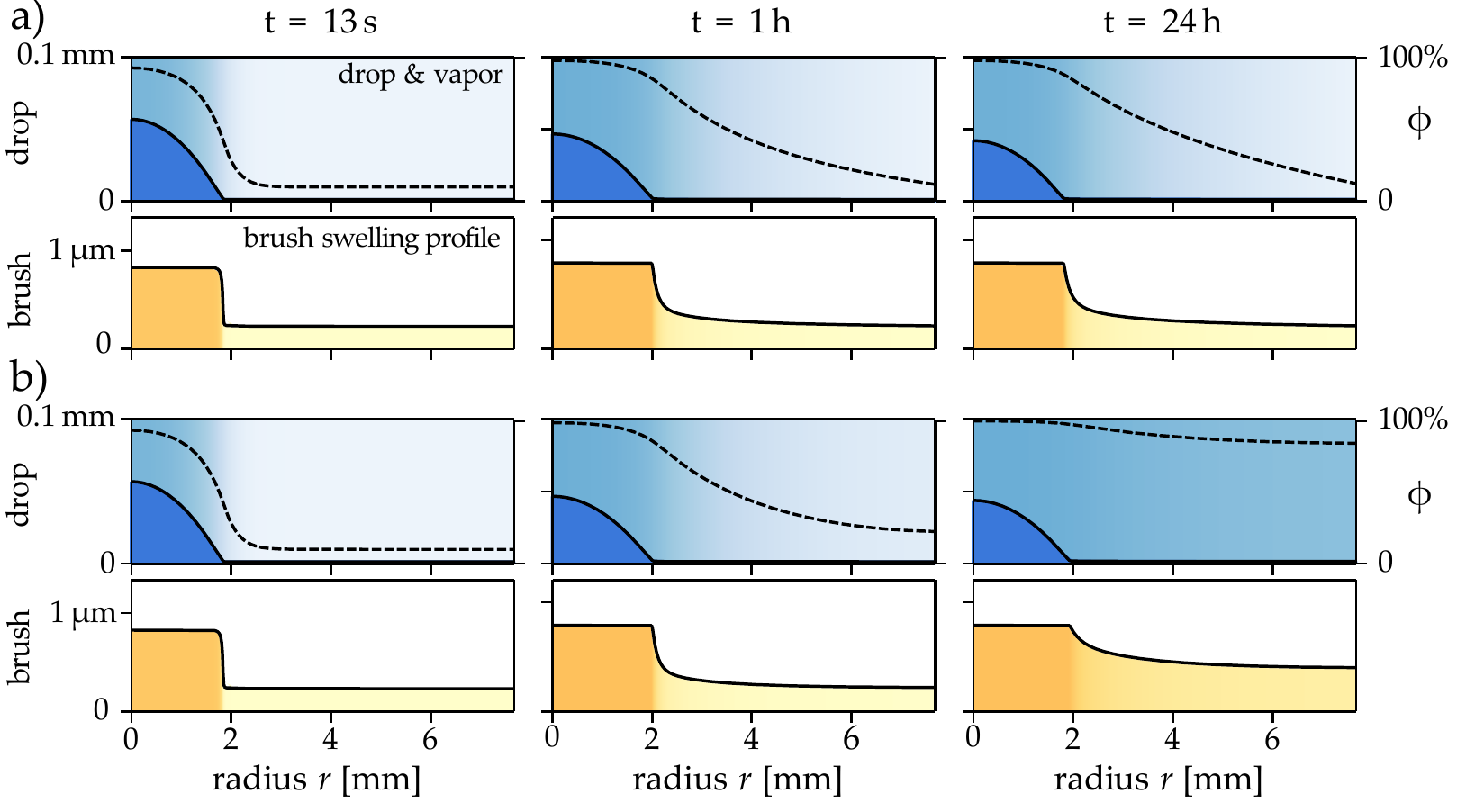}
    \caption{Shown are the results of numerical simulations for the coupled evolution of drop profile (top panels; solid lines shaded dark blue), vapor saturation profile (top panels; light blue graded shading and dashed lines), and brush swelling profile (bottom panels; graded shading in orange and solid lines) for (a) the open configuration and (b) the closed configuration. Note the different horizontal and vertical scales.}
    \label{fig:Figure 6.png}
\end{figure}

To reach a detailed understanding of the dominant transport processes and of the origin of the non-equilibrium stationary state characterized by steady profiles, we perform numerical calculations of the combined drop spreading and brush swelling process using the gradient dynamics model described in section~\ref{sec:model}. In all simulations, the drops are placed at $t=0$ on an initially dry sample in a chamber with a dry atmosphere. (For numerical reasons, we actually chose small but finite initial oil saturations of 4\,\% and 10\,\% for the brush layer and the atmosphere, respectively, rather than numerically ill-defined completely dry initial conditions.) The open configuration is implemented by imposing a constant vapor saturation of 10\% along the right edge of the simulation box, while for the closed configurations, a no-flux condition is used (see Figure~\ref{fig:Fig1.png}b). Within a fraction of the first second, oil quickly penetrates and completely saturates the brush layer directly underneath the drop (indicated by the saturated orange in the left column of Figures~\ref{fig:Figure 6.png}a and b). At the same time, the oil evaporates from the drop surface and quickly generates an almost saturated vapor phase directly above the drop (blue shading of the gas layer in the left column of Figures~\ref{fig:Figure 6.png}a and b). Diffusion subsequently allows the oil molecules to spread out in the radial direction both in the vapor phase and within the brush layer, as visualized by the softening gradient of the vapor saturation profiles in the top panels as well as of the brush saturation profiles in the bottom panels of Figures~\ref{fig:Figure 6.png}a and b. The solid lines in the latter panels correspond directly to the thickness profiles of the brush layers. 
Note that the brush model predicts the existence of a wetting ridge, as shown by Greve et al. 
\cite{GrHT2023arxiv}. The wetting ridge is too small to be visible in Figure~\ref{fig:Figure 6.png} due to our choice of parameters, namely, the strength of the brush potential.

A further observation in Figure~\ref{fig:Figure 6.png} is that after 1\,h (middle column), the open and the closed configuration show almost identical vapor saturation and brush swelling profiles. Only at a later stage (e.g., after 24\,h as shown in the right column), the vapor saturation becomes nearly uniform in the closed configuration while an almost linear vapor saturation profile develops in the open configuration. This key difference between the two configurations arises from the different boundary conditions imposed on the vapor concentration profile on the right-hand boundary. The different vapor saturation profiles are accompanied by different brush swelling profiles: In the open configuration, the profile after 24\,h is much closer to the one after 1\,h than in the closed configuration.

These results are summarized in the right column of Figure~\ref{fig:profiles}, which provides a direct comparison with the experimental profiles in the left column that we have discussed above. The model reproduces all salient features of the experimental observations, namely, the (near) stationary character of the profiles in the open configuration and the gradual evolution along with a continuous swelling far away from the contact line for the closed configuration.
Note that the absolute swelling ratios slightly differ between experiment and simulations, likely because the assumption of a fully collapsed brush in the dry limit $\zeta\to 0$ of the model is idealized.
Moreover, the decay of the stationary halo profile to a constant height in the open configuration (Figure~\ref{fig:profiles}a) is slower in the model than in the experiment. This is a consequence of the implementation of the experimental open-to-ambient-air situation via lateral boundary conditions far away from the drop in our modelling approach.

Achieving the (semi-)quantitative agreement shown in these graphs, including the absolute time scales, requires careful adjustment of several parameters in the model. The most important parameter to be fixed turns out to be the ratio between the vapor diffusion coefficient of HD, here assumed as $D_\mathrm{vap}=\SI{e-5}{m^2/s}$, and the (also diffusive) oil transport coefficient within the brush layer, $D_\mathrm{brush}$. Good agreement of the profiles is only achieved if the diffusion in the brush is chosen substantially smaller than $D_\mathrm{vap}$. The numerical results shown here correspond to $D_\mathrm{brush}=\SI{e-10}{m^2/s}$. To our knowledge, this provides a new and unique method to estimate solvent transport coefficients within a swelling polymer brush layer. Such information should be of interest whenever one considers the response time of polymer brushes to external stimuli, e.g., in sensing applications. There are, however, a few caveats. First of all, the value provided here should be considered an averaged 'effective' diffusion coefficient within the limitations of our model. The model neglects possible variations of $D_\mathrm{brush}$ with the degree of solvent saturation in the brush. Moreover, the absolute value of $D_\mathrm{vap}$ is expected to depend also on the transfer coefficients that relate the fluxes $J_\mathrm{im}$, $J_\mathrm{ev}$, and $J'_\mathrm{ev}$ to the differences between the chemical potentials of the oil in the adjacent phases. The values assumed for these quantities (see Appendix) are subject to a substantial uncertainty that has an important impact on the absolute value of $D_\mathrm{vap}$. To minimize the influence of this uncertainty, here, we assume that both diffusive processes are slower than the actual phase change, i.e., we consider a diffusion-limited case. A more detailed analysis of the absolute values would require a more extensive set of experiments to further constrain the numerical parameters. 

\begin{figure}[hbt!]
    \centering
    \includegraphics[width=\textwidth]{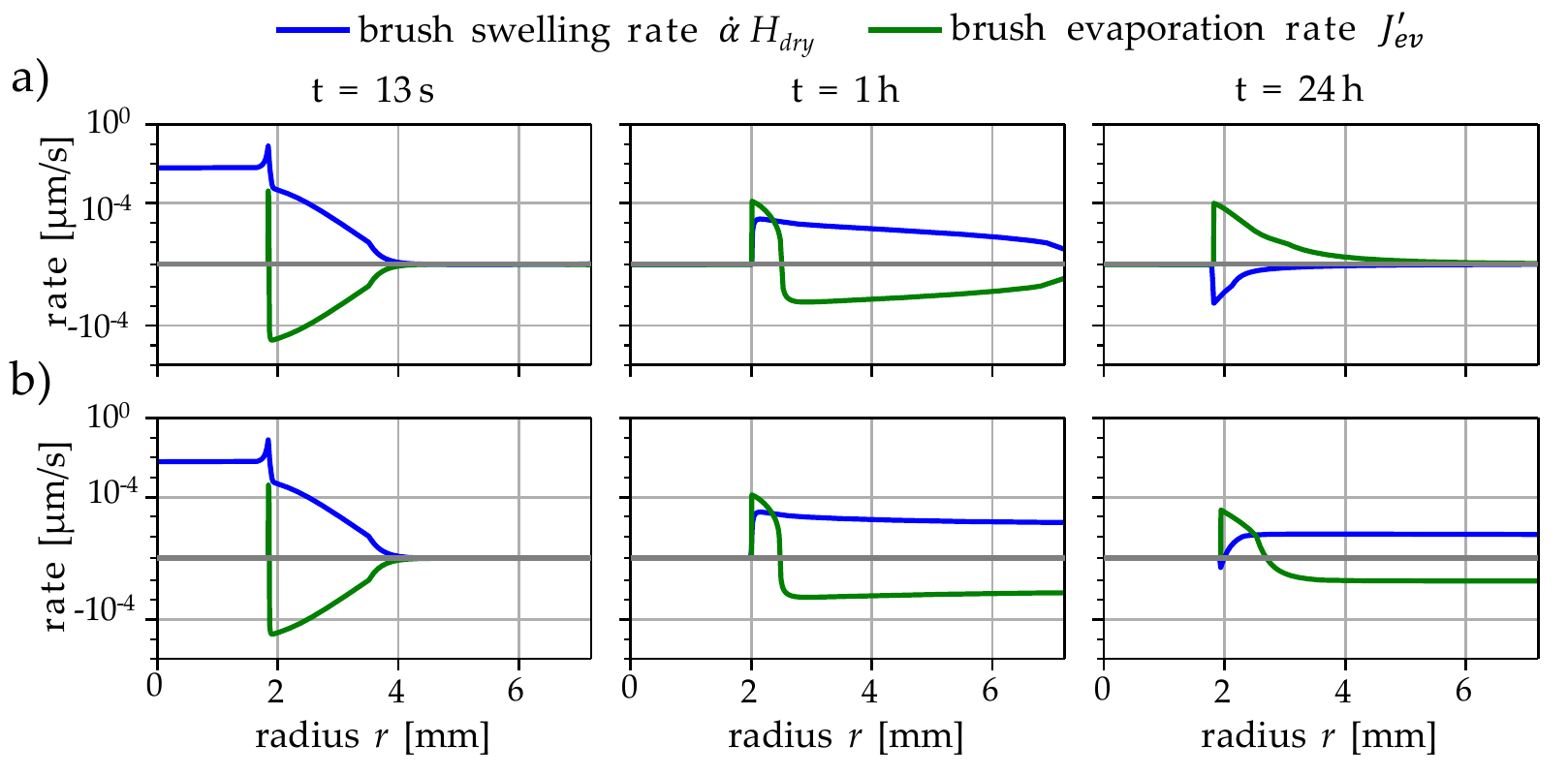}
    \caption{Numerically obtained local swelling rate of the brush $\dot \alpha H_\mathrm{dry}$ (height per time, blue lines) and the local rate of liquid evaporation from the brush $J_\mathrm{ev}'$ (liquid volume per area per time $\hat{=}$ height per time, green lines) at different instances of time $t$ corresponding to Figure~\ref{fig:Figure 6.png}. We again distinguish between (a) open configuration and (b) closed configuration. The contact line is situated at the left end of the respective green line. Note that the rates are visualized on a symmetric log axis with a linear scale between $\pm\SI{e-6}{\micro m/s}$.
    }\label{fig:fluxes}
\end{figure}

Notwithstanding these limitations, several additional conclusions can be extracted from the numerical simulations: the consequences of the faster transport in the vapor phase can be seen in Figure~\ref{fig:fluxes}. There, we show the local brush swelling rate  (blue lines) and the contribution due to evaporation from the brush layer into the vapor phase (green lines) for the simulations corresponding to the snapshots in Figure~\ref{fig:Figure 6.png}. The faster diffusion in the vapor phase leads to a quickly increasing vapor saturation in the vicinity of the contact line, while the underlying brush layer is still dry. In  consequence, the brush layer acts as a sink and swells by absorbing oil from the vapor phase. This corresponds to initially negative values of the brush evaporation rate close to the contact line (green) accompanied by the positive total brush swelling rate (blue). At later times ($t=\SI{1}{h}$), the situation has reversed: the brush layer is now fairly swollen close to the contact line. The brush layer is efficiently fed with oil by imbibition within the polymer layer. In consequence, the brush saturation exceeds the local vapor saturation and the flux from the brush into vapor becomes positive indicating net evaporation close to the contact line. Farther away from the contact line, the original situation prevails: the vapor saturation is higher than the brush saturation and brush swelling is dominated by oil condensation from the vapor. At very late stages ($t=\SI{24}{h}$), clear differences in the fluxes appear between the open and the closed configuration. As one might expect, for the open configuration the low vapor saturation far away from the contact line leads to continuous evaporation of oil from the brush layer. This explains the existence of the non-equilibrium stationary state related to steady swelling profiles: they result from the balance between continuous evaporation and continuous influx of oil by imbibition within the brush layer. This continuous flux stabilizes the prevailing gradients in brush layer thickness characterizing the stationary state. A simplified version of a similar mechanism was in fact already proposed by Seker et al. \cite{seker2008kinetics} to explain imbibition of volatile fluids into a porous medium that is surrounded by a dry atmosphere. For the closed configuration in our experiments, far away from the contact line, net condensation dominates even at very large times as the vapor approaches full saturation more quickly than the brush layer. The fact that after 24\,h the brush layer still displays a substantial thickness gradient despite the high saturation is due to the fact that the vapor phase is still not completely saturated at the right-hand side of our simulation box. Given the fact that the equilibrium adsorption isotherm of our system is very steep upon approaching complete saturation, even a minor undersaturation of 5--10\,\% still leads to a substantial reduction of the brush layer thickness. 

Finally, it is worthwhile to comment on the observed very long relaxation times and the fact that even in the closed configuration, the system still evolves after 24\,h. At first glance, this may seem surprising given the fact that the characteristic time scale for vapor diffusion in the system is 10\,s. Because of the combination of the low absolute vapor pressure of HD and the high sorption capacity of the brush layer, the transient states in our system display a substantially larger lifetime. While the diffusion time is indeed of the order of a few seconds, transporting the equivalent of a film of a few hundred nanometers height of liquid HD as required to saturate the brush layer takes much longer: a simple estimate yields an equilibration time for the system of $T_\mathrm{eq}=\rho_\mathrm{liq}L\Delta \zeta / D_\mathrm{vap} \rho_\mathrm{vap} \approx \SI{2.4e5}{s}$, which is of the order of days. This is consistent with the observation that after 24\,h the brush layer is still far from being homogeneously swollen. From this expression, we see that equilibration should accelerate with increasing vapor pressure, as intuitively plausible. Preliminary experiments with drops of tetradecane and dodecane with vapor pressures at room temperature of \SI{1.55}{Pa} and \SI{18}{Pa}, respectively, instead of \SI{0.2}{Pa} for hexadecane confirm this expectation (data not shown). For water drops with a vapor pressure of \SI{2300}{Pa} spreading on swellable responsive surface coatings, including polymer brush and polyelectrolyte layers, the influence of vapor phase transport should be even more important.

\section{Conclusions}

In summary, we have demonstrated that the spreading of drops of volatile hexadecane on hydrophobic polymer brush layers of PLMA is accompanied by the formation of a halo of partly swollen brushes. Swelling kinetics and the extent of the halo are controlled by the balance of two competing transport mechanisms, namely, direct imbibition of oil from the drop through the polymer brush layer and vapor phase transport in combination with evaporation and condensation at the brush-vapor interface. Numerical simulations with a mesoscopic hydrodynamic model based on a gradient dynamics framework reproduce the experimentally observed time-dependent swelling profiles for slowly evaporating drops in both an open atmosphere and in a closed cell. Matching the numerical results to the experimental data provides a method to estimate the hitherto unknown diffusion coefficient of the solvent within the polymer brush layer, which for the present system is found to be approximately \num{10000} times lower than the diffusion coefficient in vapor. The combination of this small diffusion coefficient and the low vapor pressure explains the very long relaxation times of more than 24\,h. We anticipate that vapor phase transport should play an important role in many dynamic wetting phenomena on swellable polymer materials and coatings, in particular for aqueous drops with their characteristic high vapor pressure. Our experiments also suggest that the strong gradients in the local swelling of such responsive systems can be achieved by regulating the local vapor saturation in a controlled manner. This may be of interest to sensing applications. 

\section*{Acknowledgements}
SdB, UT and SH acknowledge support by the Deutsche Forschungsgemeinschaft (DFG) within SPP~2171 by Grants No.\ BE8029/1-2, TH781/12-1 and TH781/12-2.

\newpage 
\appendix
\section{Full theoretical model}
\subsubsection{Three-field gradient dynamics}

The dynamics of thin films or drops of a nonvolatile liquid on solid substrates are often described by reduced models for the evolution of the film thickness profile. These thin-film (or long-wave) models are obtained through a long-wave approximation~\cite{OrDB1997rmp,CrMa2009rmp} and can often be written in a gradient dynamics form~\cite{Mitl1993jcis, Thie2018csa}. This form accounts on the one hand for convective transport processes through mass-conserving terms in the form of a conservation law, so-called ``conserved contributions''. On the other hand, the form accounts for condensation/evaporation by nonconserved contributions. In the limiting case of mass transfer-limited phase change~\cite{LyGP2002pre,Thie2014acis}, the resulting kinetic equation for the height profile $h(\mathbf x, t)$ in compact gradient dynamics form is~\cite{Thie2010jpcm}
\begin{equation}
    \partial_t h(\mathbf x, t) \,=\,
    \nabla\cdot\left[Q(h)\,\nabla\frac{\delta
            \mathcal{F}}{\delta h}\right]
    - M(h)\,\left(\frac{\delta \mathcal{F}}{\delta h} -p_\mathrm{vap}\right).
    \label{eq:onefield:gov}
\end{equation}
Here, $\mathbf x=(x,y)^T$ are the substrate coordinates, and the expressions \(Q(h)\ge0\) and \(M(h)\ge0\) are the mobility functions for the conserved and the nonconserved part of the dynamics, respectively. All parts of the dynamics are driven by the free energy functional \(\mathcal{F}[h]\) incorporating, e.g., the liquid-gas interface energy, wetting energy and potential energy.
Additionally, a constant external vapor pressure \(p_\mathrm{vap}\) is imposed, and controls the flux due to phase change.

In more complex systems, the dynamics of the drop/film profile couples to other dynamic quantities characterizing the system. Then, the model can be extended to a gradient dynamics of multiple coupled order parameter fields, e.g., effective solvent and solute height profiles for films/drops of mixtures of simple liquids~\cite{ThTL2013prl,XuTQ2015jpcm} or drop profile and surfactant concentration profiles in and on the liquid for films/drops covered by a soluble surfactant~\cite{ThAP2016prf,Thie2018csa}.

Here, we aim at a description of the coupled dynamics of drop/film, brush-contained liquid, and vapor density profiles. Hence, we consider a gradient dynamics model for the case of three variables $\psi_a(\mathbf x, t)$ in the general form \cite{Thie2018csa}
\begin{equation}
    \partial_t \psi_a=\nabla\cdot  \left[ \sum_{b=1}^3 \,Q_{ab}\,
    \nabla \frac{\delta \mathcal{F}}{\delta \psi_b}\right]
    - \sum_{b=1}^3 \,M_{ab}\, \frac{\delta \mathcal{F}}{\delta \psi_b},
    \label{eq:three-field-gradient-dynamics}
\end{equation}
where the subscripts \(a, b = 1,2,3\) refer to the three fields and the $3\times 3$ mobility matrices $M_{ab}, Q_{ab}\neq0$ represent the conserved and nonconserved dynamics, respectively.

\begin{figure}
    \centering
    \includegraphics[]{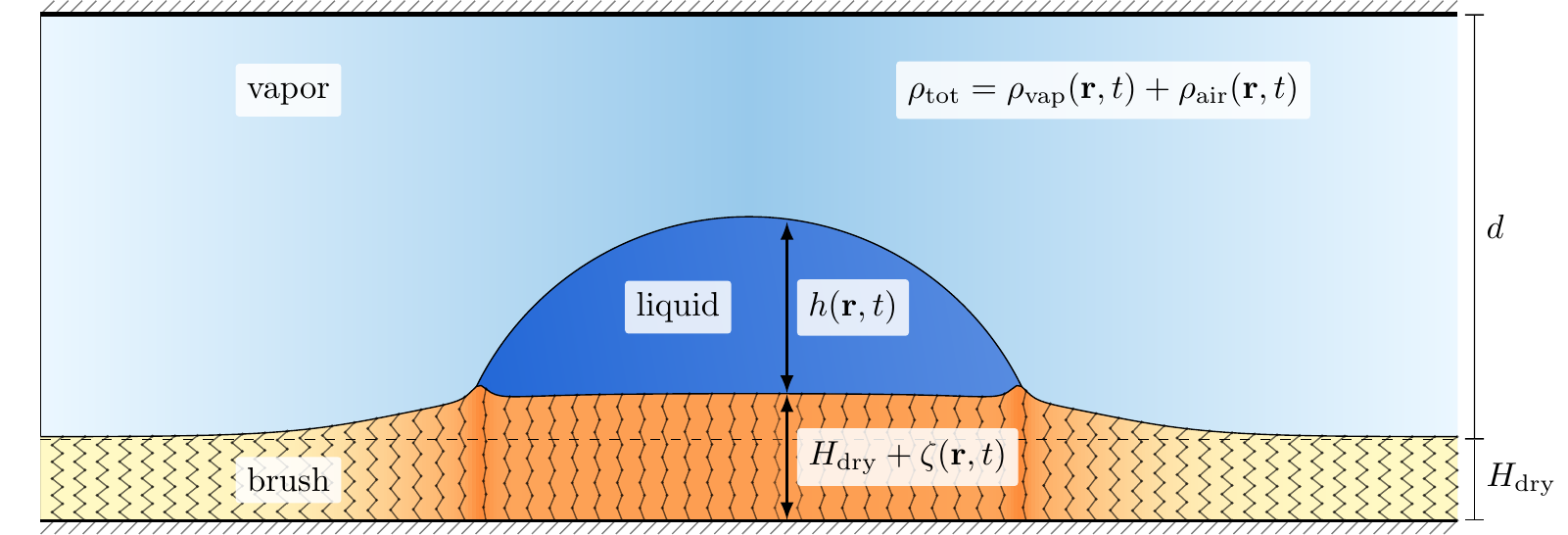}
    \caption{Sketch of the considered geometry for a volatile liquid drop on a polymer brush within a chamber of height $d+H_\mathrm{dry}$. The drop profile is described by the height $h(\mathbf x, t)$ and the brush height consists of its dry height $H_\mathrm{dry}$ and the effective height of the imbibed liquid $\zeta(\mathbf x, t)$. The particle densities of vapor $\rho_\mathrm{vap}(\mathbf x, t)$ and ambient air $\rho_\mathrm{air}(\mathbf x, t)$ together account for a constant total density $\rho_\mathrm{tot}$ in the gas phase. The dimensions are illustrative and not scaled.}\label{fig:brush-sketch}
\end{figure}

For the description of the coupled brush and liquid dynamics, we adapt the approach developed in Ref.~\cite{ThHa2020epjst}, where the brush state is solely characterized by the local amount of imbibed liquid, i.e., the local effective height of the liquid in the brush $\zeta(\mathbf x, t)$, as illustrated in Figure~\ref{fig:brush-sketch}. We thereby approximate the state of the brush as vertically homogeneous. As we assume that any height increase of the brush is solely caused by the imbibed liquid, the effective height \(\zeta\) directly relates to the swelling ratio $\alpha$ of the brush:
\begin{equation}
    \alpha(\mathbf x, t) = \frac{H_\mathrm{wet}(\mathbf x, t)}{H_\mathrm{dry}} =  \frac{H_\mathrm{dry} + \zeta(\mathbf x, t)}{H_\mathrm{dry}}\label{eq:swelling_ratio}
\end{equation}
where $H_\mathrm{dry}$ denotes the `dry' height of the brush, that is, the brush height in an unswollen state ($\alpha = 1$).

Since the vapor is confined to a narrow gap between the drop/film and the top closure of the chamber of a large aspect ratio, we also assume that the vapor distribution is approximately homogeneous in the vertical direction. Adapting the approach of Ref.~\cite{HDJT2023jfm} this allows us to describe the vapor particle density in the chamber by a single field $\rho_\mathrm{vap}(\mathbf x, t)$ that does not depend on the vertical coordinate $z$ (cf.~Figure~\ref{fig:brush-sketch}). In this way, the vapor concentration serves as the third field variable, effectively characterizing the local amount of vapor in the gap above the brush and the drop/film. In the following, we consider air and vapor as ideal gases. Thus, the vapor particle density $\rho_\mathrm{vap}$ relates to the vapor saturation $\phi$ as
\begin{equation}
    \phi(\mathbf x, t) = \frac{\rho_\mathrm{vap}(\mathbf x, t)\,k_B T}{p_\mathrm{sat}}
\end{equation}
where $p_\mathrm{sat}$ is the constant saturation pressure  of the liquid.

For a thermodynamically sensible description in the gradient dynamics framework~\eqref{eq:three-field-gradient-dynamics}, we first transform all three order parameter fields to particle numbers per area, i.e.\ the per area number of liquid molecules in the drop $\psi_1(\mathbf x, t) = \rho_\mathrm{liq}\,h(\mathbf x, t)$, within the brush $\psi_2(\mathbf x, t) = \rho_\mathrm{liq}\,\zeta(\mathbf x, t)$, and in the ambient air $\psi_3(\mathbf x, t) = \rho_\mathrm{vap}(\mathbf x, t)\,[d-h(\mathbf x, t)-\zeta(\mathbf x, t)]$. Here, we have introduced the vapor particle density $\rho_\mathrm{vap}(\mathrm x, t)$ and the constant liquid particle density $\rho_\mathrm{liq}$. Conveniently, all variations of the free energy functional with respect to the particle number densities $\psi_a$ then correspond to effective chemical potentials $\mu_a = \delta F / \delta \psi_a$.

\subsubsection{Transport processes}
Next, we provide expressions for the mobility matrices $\mathbf{\underline{Q}}$ and $\mathbf{\underline{M}}$ in Eq.~\eqref{eq:three-field-gradient-dynamics} by considering the transport processes in the system. We call all terms containing $\mathbf{\underline{Q}}$ in Eq.~\eqref{eq:three-field-gradient-dynamics} ``conserved'', as they define lateral particle fluxes within the respective region (brush, drop, and vapor).
Following the approach of Refs.~\cite{ThHa2020epjst,HDJT2023jfm}, the conserved dynamics only accounts for three processes: (i) viscous motion within the drop, (ii) diffusive transport of liquid particles within the brush, and (iii) diffusive transport of vapor particles within the vapor. In this way we neglect dynamic coupling between the regions, e.g.\ we assume there is no viscous drag across the boundary between drop and brush.
This results in the diagonal matrix
\begin{equation}
    \mathbf{Q} = \begin{pmatrix}
        \frac{1}{\rho_\mathrm{liq}}\,\frac{\psi_1^3}{3 \eta} & 0 & 0\\
        0 & \frac{1}{k_BT} D_\mathrm{brush} \psi_2 & 0\\
        0 & 0 & \frac{1}{k_BT} D_\mathrm{vap} \psi_3
    \end{pmatrix},
    \label{eq:conserved-mobility}
\end{equation}
where we have introduced the liquid dynamic viscosity $\eta$ and the diffusion coefficients $D_\mathrm{vap}$ and $D_\mathrm{brush}$ of the vapor in the air and of the liquid in the brush, respectively.

Accordingly, any transport via the nonconserved part of the dynamics corresponds to transfer processes of particles from one region to the other, e.g., from drop to brush and from drop to vapor. For the sake of simplicity, here, we assume that any such transfer is directly proportional to a difference in the corresponding chemical potentials. In particular, this implies that there is no direct dependency of the transfer rate on the fields $\psi_a$. We then explicitly incorporate transfer processes between all phases and respective transfer rate (Onsager) coefficients, namely (i) between drop and brush (imbibition) via the coefficient $M_\mathrm{im}$, (ii) between drop and vapor (evaporation/condensation) via the coefficient $M_\mathrm{ev}$, and (iii) directly between the brush and vapor (evaporation/condensation) via a coefficient $M_\mathrm{ev}'$. The resulting nonconserved mobility matrix is
\begin{equation}
    \mathbf{M} = \begin{pmatrix}
        M_\mathrm{im} + M_\mathrm{ev} & -M_\mathrm{im}                 & -M_\mathrm{ev}\\
        -M_\mathrm{im}                & M_\mathrm{im} + M_\mathrm{ev}' & -M_\mathrm{ev}'\\
        -M_\mathrm{ev}                & -M_\mathrm{ev}'                & M_\mathrm{ev}+M_\mathrm{ev}'
    \end{pmatrix}.
    \label{eq:nonconserved-mobility}
\end{equation}
Notably, the symmetry of the matrix reflects the fact that all transfer processes are allowed in both directions. Furthermore, as the total number of particles is locally conserved, the sum of the three fields (the total particle number per area) fulfils a continuity equation $\partial_t (\psi_1 + \psi_2 + \psi_3) = -\nabla\cdot\mathbf j$ with the total flux $\mathbf j$. In other words, each row of $\mathbf{M}$ [\eqref{eq:nonconserved-mobility}] adds up to zero.

Note that the given description of the transport processes includes some unwanted side effects. In particular, it allows for evaporation (and condensation) of liquid from (to) the brush in areas that are covered by the drop. This can be fixed by modulating the respective transfer coefficient $M_\mathrm{ev}' = M_\mathrm{ev}'(h)$ with a smooth step function such that it is close to zero when the drop profile height $h(\mathbf x, t)$ is larger than a small threshold value and otherwise constant. As our model incorporates a thin liquid adsorption layer to avoid the contact line singularity~\cite{HuSc1971jcis}, we choose the threshold height slightly larger than the equilibrium adsorption layer height. Similarly, we modulate the two transfer coefficients $M_\mathrm{ev}$, $M_\mathrm{im}$ in order to suppress any imbibition or evaporation of liquid from the film when the profile height is smaller than the threshold value.
This is necessary mostly for two reasons: First, if the adsorption layer was coupled to the vapor or to the brush, the height of the adsorbed film would increase slightly such that the pressures in film, vapor and brush balance. While this effect may be very subtle, it can take up a substantial amount of liquid across a large domain, effectively draining the drop as the adsorption layer adapts to changes in the atmosphere or brush state. Second, gradients in the brush or vapor pressures would also evoke a gradient in the film pressure, hence causing a liquid flux through the adsorption layer. In this way, the model would bypass the `slow' diffusive transport processes by rapidly transferring liquid away from the droplet via the adsorption layer, where it then evaporates or absorbs.
As an alternative to the modulation of the transfer coefficients described above, both effects could also be suppressed by employing an ultra-thin adsorption layer height, which would on the other hand inhibit contact line motion.

\subsubsection{Energy functional}\label{sec:en}
Having established a simple dynamical framework, next we discuss the underlying free energy functional $F[\psi_1,\psi_2,\psi_3]$ that determines all the chemical potentials $\mu_i = \delta F/\delta \psi_i$ driving the dynamics.

First, we write it in terms of \(h\), \(\zeta \), and \(\rho_\mathrm{vap}\) as
\begin{align}
    F = \int_\Omega \Bigg[
        \underbrace{\gamma_\mathrm{lg} {\sqrt{1+|\nabla (h+\zeta)|^2}}}_\text{liquid-gas\,interface\,energy}
        + \underbrace{\gamma_\mathrm{bl}(\zeta)\sqrt{1+|\nabla \zeta|^2}}_\text{liquid-brush\,interface\,energy}
        + \underbrace{f_\mathrm{wet}(h,\zeta)\sqrt{1+|\nabla \zeta|^2}}_\text{wetting potential}\nonumber\\
        + \underbrace{f_\mathrm{brush}(\zeta)}_\text{brush energy}
        + \underbrace{(h+\zeta) f_\mathrm{liq}(\rho_\mathrm{liq})}_\text{liquid bulk energy}
        + \underbrace{(d-h-\zeta) f_\mathrm{vap}(\rho_\mathrm{vap})}_\text{vapour energy}
        + \underbrace{(d-h-\zeta) f_\mathrm{air}(\rho_\mathrm{air})}_\text{air energy}
        \Bigg] \mathrm{d}^2x,\label{eq:free-energy-functional}
\end{align}
where \(f_\mathrm{liq}\), \(f_\mathrm{vap}\), and \(f_\mathrm{air}\) are bulk liquid, vapor, and air energies per volume, that we convert to per area energies by multiplying them with the effective liquid height and gap height, respectively. Note that the energy for the liquid state applies to the liquid within drop and brush, i.e.,  \(f_\mathrm{liq}\) has to be multiplied by $h+\zeta$. Furthermore, \(\gamma_\mathrm{lg} \) is the constant liquid-gas interface energy, \(\gamma_\mathrm{bl}\) is the brush state-dependent effective liquid-brush interface energy, and \(f_\mathrm{wet}(h,\zeta)\) is the brush state-dependent (per area) wetting energy. The interface energies are each multiplied by a corresponding metric factor accounting for the local interface length. 

We employ a wetting potential that allows for partial wetting, i.e., a potential that accounts for a finite equilibrium contact angle. Typically, the wettability of a polymer brush depends on the liquid content~\cite{GuGu2016ra, BBSV2018l}, i.e., the swelling ratio $\alpha$. Here, we adapt the magnitude of the potential, that is directly related to the equilibrium contact angle, via a simple power law:
\begin{align}
    f_\mathrm{wet}(h, \zeta) = \left(-\frac{A}{2h^2} + \frac{B}{5h^5}\right) \frac{1}{\alpha^{\beta}}\label{eq:wetting_potential}
\end{align}
with some positive exponent \(\beta > 0\). In other words, the wettability of the brush increases as the liquid content of the brush increases.
In Refs.~\cite{Bonn2001cocis, Inde2010pa} similar power laws were observed for a thermal adaption of the wetting properties of (non-polymeric) surfaces.
The Hamaker-type constants $A$ and $B$ represent the respective strengths of short- and long-range forces. In consequence, they determine the height of the adsorption layer $h_p = \sqrt[3]{B/A}$ that covers the substrate in macroscopically dry (non-wetted) regions.
Moreover, the wetting energy defines the equilibrium contact angle $\theta_e$ via $\cos \theta_e = 1 + f_\mathrm{wet}(h_p) / \gamma_\mathrm{lg}$~\cite{Chur1995acis,TSTJ2018l}, which for the adaptive wetting potential~\eqref{eq:wetting_potential} implies
\begin{equation}
    \theta_e \approx \sqrt{\frac{3A}{5\,\alpha^\beta\,\gamma_\mathrm{lg}\,h_p^2}},
\end{equation}
i.e., as intended $\theta_e$ decreases with increasing swelling and approaches zero with diverging liquid content. Note that this simple ansatz may be adapted to more intricate dependencies of wettability on brush state.

Similarly, we assume an adaption of the brush-liquid interface energy $\gamma_\mathrm{bl}$ to the brush state. We employ a power law with the same power:
\begin{equation}
    \gamma_\mathrm{bl}(\zeta) = \frac{\gamma_\mathrm{bl,0}}{\alpha^\beta},
\end{equation}
thereby assuming that $\gamma_\mathrm{bl}$ decreases as the brush swells and also approaches zero with diverging liquid content. Here, the constant $\gamma_\mathrm{bl,0}$ denotes the surface energy of the dry brush. We acknowledge, that this ansatz may not be the most general, yet it ensures consistency between mesoscopic and macroscopic descriptions of the three-phase contact region, as, e.g., discussed for the case of a droplet covered by a soluble surfactant by \cite{TSTJ2018l}.

Next, we specify the brush energy. As widely found in the literature~\cite{Doi2013, Somm2017m, ritsema2020sorption, ThHa2020epjst}, the free energy of the brush-solvent system includes an elastic contribution from the stretching polymers and entropic contributions described by the Flory-Huggins model. Using the Kuhn length $\ell_K$, i.e., the length of a unit cell in the lattice model, and the relative grafting density $\tilde \sigma = \sigma \ell_K^2$ (number of polymers grafted per unit area) we write for the per area brush energy
\begin{equation}
    f_\mathrm{brush}(\zeta) = \frac{H_\mathrm{dry} k_B T}{\ell_K^3}\left[\frac{\tilde \sigma^2}{2} \alpha^2 +
        (\alpha - 1) \,\log\left(1-\frac{1}{\alpha}\right) + \chi \left(1-\frac{1}{\alpha}\right)\right],
    \label{eq:brush-en-area}
\end{equation}
as obtained by integrating the per volume free energy over the brush height $\alpha H_\mathrm{dry}$. Note that the factor $1/\ell_K^3$ relates to a density in the Flory-Huggins lattice model that is for simplicity commonly equated with the liquid density $\rho_\mathrm{liq}$. For more details on the modelling of the drop-brush subsystem see Ref.~\cite{ThHa2020epjst}.

If vapor and air are considered ideal gases, we can directly give their respective free energy densities as
\begin{align}
    f_\mathrm{vap} = k_B T \rho_\mathrm{vap} \left[ \log(\Lambda^3\rho_\mathrm{vap}) - 1 \right]
    ~~\text{and}~~
    f_\mathrm{air} = k_B T \rho_\mathrm{air} \left[ \log(\Lambda^3\rho_\mathrm{air}) - 1 \right]
\end{align}
with the constant total density $\rho_\mathrm{tot} = \rho_\mathrm{air}(\mathbf x, t) + \rho_\mathrm{vap}(\mathbf x, t)$ and the mean free path length $\Lambda$, which cancels immediately in Eq.~\eqref{eq:free-energy-functional}.

Considering the equilibrium of a thick liquid film in an atmosphere saturated with vapor reveals a relation between the saturation vapor pressure $p_\mathrm{sat}=\rho_\mathrm{sat}k_BT$ and the bulk liquid free energy $f_\mathrm{liq}$, which we use to determine the value of the latter as
\begin{equation}
    f_\mathrm{liq} = \rho_\mathrm{liq} k_B T \log \left( \frac{\rho_\mathrm{sat}}{\rho_\mathrm{tot} - \rho_\mathrm{sat}} \right).\label{eq:f_liq}
\end{equation}
For more details on the modelling of the drop-vapor subsystem see Ref.~\cite{HDJT2023jfm}.

\subsubsection{Resulting model equations}
To obtain the explicit form of the dynamical equations, we evaluate the variations of the free energy with respect to the three fields. The resulting chemical potentials are
\begin{equation}
    \begin{aligned}
        \mu_\mathrm{film} = \frac{\delta F}{\delta \psi_1}  & = \frac{1}{\rho_\mathrm{liq}} \left[ -\gamma_\mathrm{lg} \frac{\Delta (h+\zeta)}{\xi_{h+\zeta}^3} + \xi_\zeta \partial_h f_\mathrm{wet}(h, \zeta) + f_\mathrm{liq} \right],                                         \\
        \mu_\mathrm{brush} = \frac{\delta F}{\delta \psi_2} & = \frac{1}{\rho_\mathrm{liq}} \bigg[ -\gamma_\mathrm{lg} \frac{\Delta (h+\zeta)}{\xi_{h+\zeta}^3} - \nabla \left\{ \left[ \gamma_\mathrm{bl}(\zeta) + f_\mathrm{wet}(h, \zeta) \right] \frac{\nabla \zeta}{\xi_\zeta} \right\}\\
            & \hspace{3em}+ \xi_\zeta \partial_\zeta \left[ \gamma_\mathrm{bl}(\zeta) + f_\mathrm{wet}(h, \zeta) \right] + \partial_\zeta f_\mathrm{brush}(\zeta) + f_\mathrm{liq} \bigg], \\
        \mu_\mathrm{vap} = \frac{\delta F}{\delta \psi_3}   & = k_B T \log \left( \frac{\rho_\mathrm{vap}}{\rho_\mathrm{tot} - \rho_\mathrm{vap}} \right),
    \end{aligned}\label{eq:chem_potentials}
\end{equation}
where we have utilized that the vapor particle density is much smaller than the total gas particle density, which itself is much smaller than the liquid density $\rho_\mathrm{vap} \ll \rho_\mathrm{tot} \ll \rho_\mathrm{liq}$, to simplify the expressions. The metric factors are abbreviated as $\xi_{h+\zeta}=\sqrt{1+|\nabla (h+\zeta)|^2}$ and $\xi_{\zeta}=\sqrt{1+|\nabla \zeta|^2}$.

Inserting the obtained variations into the three-field gradient dynamics\ \eqref{eq:three-field-gradient-dynamics} gives the kinetic equations:
\begin{equation}
    \begin{aligned}
        \partial_t \psi_1 / \rho_\mathrm{liq}  & = \nabla\cdot \left[ \frac{\psi_3^3}{3\eta \, \rho_\mathrm{liq}^2} \nabla \mu_\mathrm{film} \right] - J_\mathrm{ev} - J_\mathrm{im},   \\
        \partial_t \psi_2  / \rho_\mathrm{liq} & = \nabla\cdot \left[ \frac{D_\mathrm{brush}}{\rho_\mathrm{liq} k_B T} \, \psi_2 \, \nabla \mu_\mathrm{brush} \right] - J_\mathrm{ev}' + J_\mathrm{im},                 \\
        \partial_t \psi_3 / \rho_\mathrm{liq}  & = \nabla\cdot \left[ D_\mathrm{vap} (d-h-\zeta) \nabla \rho_\mathrm{vap} \right] / \rho_\mathrm{liq} + J_\mathrm{ev} + J_\mathrm{ev}',
    \end{aligned}\label{eq:dynamic_eqs1}
\end{equation}
where the fluxes transporting particles between the regions are
\begin{align}
    J_\mathrm{ev} = \frac{M_\mathrm{ev}}{\rho_\mathrm{liq}} (\mu_\mathrm{film} - \mu_\mathrm{vap}),\quad
    J_\mathrm{ev}' = \frac{M_\mathrm{ev}'}{\rho_\mathrm{liq}} (\mu_\mathrm{brush} - \mu_\mathrm{vap}),\quad
    J_\mathrm{im} = \frac{M_\mathrm{im}}{\rho_\mathrm{liq}} (\mu_\mathrm{film} - \mu_\mathrm{brush}),
\end{align}
namely, the drop/film evaporation/condensation flux, the evaporation/condensation flux of the liquid contained in the brush and the imbibition flux, respectively. They are all given in units of liquid volume per time and area.

Lastly, we further simplify the dynamical equations \eqref{eq:dynamic_eqs1} by exploiting that the brush is much thinner than the chamber height $\zeta \ll d$ and by substituting the particle per area variables $\psi_{1}, \psi_2$, and $\psi_3$ by the more intuitive film height $h$, the dimensionless swelling ratio $\alpha$, and vapor saturation $\phi$, respectively.
Then, the final model is
\begin{equation}
    \begin{aligned}
        \partial_t h           & = \nabla\cdot \left[ \frac{h^3\rho_\mathrm{liq}}{3\eta} \nabla \mu_\mathrm{liq} \right] - J_\mathrm{ev} - J_\mathrm{im}                                                \\
        \partial_t \alpha      & = \nabla\cdot \left[ \frac{D_\mathrm{brush}}{k_BT} \, (\alpha-1) \, \nabla \mu_\mathrm{brush} \right] + \frac{1}{H_\mathrm{dry}}(J_\mathrm{im} - J_\mathrm{ev}') \\
        \partial_t [(d-h)\phi] & = \nabla\cdot \left[D_\mathrm{vap} (d-h) \nabla \phi \right] + \frac{\rho_\mathrm{liq} k_B T}{p_\mathrm{sat}}(J_\mathrm{ev} + J_\mathrm{ev}').
    \end{aligned}\label{eq:dynamic_eqs_final}
\end{equation}

In the following, we perform time simulations of these equations using the finite-element element method implemented in the C++ library \textsc{oomph-lib}~\cite{HeHa2006}. Moreover, we make use of polar coordinates and perform all simulations for a radially symmetric geometry, effectively reducing the spatially two-dimensional cartesian domain to a one-dimensional radial domain.

\subsubsection{Model parameters}\label{sec:appendix:parameters}
The model parameters used to generate the simulation data in Figures~\ref{fig:profiles} and~\ref{fig:Figure 6.png} are given in Table~\ref{tab:params}. The parameters are chosen such that they closely match the experiments.

\begin{table}[]
\begin{tabular}{lcc}
\textbf{Parameter description} & ~\textbf{Symbol}~ & \textbf{Value} \\
\hline
viscosity & $\eta$ & \SI{3}{\milli Pa\,s} \\
ideal contact angle (dry brush) & $\theta_e$ & \SI{5}{\degree} \\
precursor layer height & $h_p$ & \SI{1}{\micro m} \\
liquid particle density & $\rho_\mathrm{liq}$ & $\frac{\SI{770}{kg / m^3}}{\SI{226}{g / mol}} N_A$ \\
vapor saturation pressure & $p_\mathrm{sat}$ & \SI{0.2}{Pa} \\
temperature & $T$ & \SI{22}{\celsius} \\
initial drop volume & $V_0$ & \SI{0.3}{\micro l} \\
initial vapor concentration & $\phi_\mathrm{lab}$ & \SI{10}{\percent}\\
liquid-gas interface energy & $\gamma_\mathrm{lg}$ & \SI{27}{\milli N / m} \\
brush-liquid interface energy (dry brush) & $\gamma_\mathrm{bl,0}$ & \SI{3}{\milli N / m} \\
relative grafting density & $\tilde \sigma$ & \SI{0.1}{} \\
dry brush height & $H_\mathrm{dry}$ & \SI{200}{nm} \\
brush lattice cell density & $1/\ell_K^3$ & $\rho_\mathrm{liq}$ \\
Flory-Huggins interaction parameter & $\chi$ & \SI{0}{} \\
brush adaption exponent (power law) & $\beta$ & \SI{1}{}\\
vapor diffusion coefficient & $D_\mathrm{vap}$ & \SI{e-5}{m^2 / s} \\
brush contained liquid diffusion coefficient & $D_\mathrm{brush}$ & \SI{e-10}{m^2 / s} \\
imbibition rate coefficient & $M_\mathrm{im}$ & \SI{e-13}{m / Pa\,s} \\
bulk liquid evaporation rate coefficient & $M_\mathrm{ev}$ & \SI{e-16}{m / Pa\,s} \\
brush contained liquid evap. rate coefficient & $M_\mathrm{ev}'$ & \SI{e-16}{m / Pa\,s} \\
simulation domain height & $d$ & \SI{1}{mm} \\
simulation domain width & $L$ & \SI{8}{mm}
\end{tabular}
\caption{Model parameters used for generating the simulation data.}
\label{tab:params}
\end{table}

Using these parameters, we next provide an estimate of the magnitude of the contributions to the local free energy as given in Eq.~\eqref{eq:free-energy-functional}. An upper bound to the local liquid-gas interface energy contribution (first term in Eq.~\eqref{eq:free-energy-functional}) for a droplet with a contact angle of $\theta_e=\SI{5}{\degree}$ is given by
\begin{equation}
    \gamma_\mathrm{lg} \xi_{h+\zeta} \approx \gamma_\mathrm{lg} = \SI{27}{mJ/m^2}.
\end{equation}
The liquid-brush interface energy contribution is even smaller, as $\gamma_\mathrm{bl}<\gamma_\mathrm{lg}$. The magnitude of the wetting potential $f_\mathrm{wet}$ can be expected to be comparable to the value of the liquid-gas interface energy, as the mesoscopic Young relation $\cos \theta_e = 1-f_\mathrm{wet}(h_p)/\gamma_\mathrm{lg}$ must hold~\cite{Chur1995acis}.

The scale of the brush energy [Eq.~\eqref{eq:brush-en-area}] is dominated by its prefactor $H_\mathrm{dry} k_BT/\ell_K^3$ while the term in brackets is roughly of the magnitude $-0.8$ for a saturated brush. For the employed parameters, the brush therefore contributes to the per area free energy with $f_\mathrm{brush} \approx -\SI{1.3}{J/m^2}$, i.e., the brush energy is much larger than the interface energies. Hence, we conclude that the intake of liquid into the brush is strongly driven by the brush potential rather than by the capillary energy of the drop.

Using Eq.~\eqref{eq:f_liq} and our assumption that $\rho_\mathrm{liq} = 1/\ell_K^3$, we can easily relate the bulk energy of the liquid contained in the brush to the brush energy scale $H_\mathrm{dry} k_BT/\ell_K^3$ as
\begin{align}
    \zeta f_\mathrm{liq} &= (\alpha - 1) H_\mathrm{dry} \rho_\mathrm{liq} k_B T \log\left( \frac{\rho_\mathrm{sat}}{\rho_\mathrm{tot} - \rho_\mathrm{sat}} \right)\\\nonumber &= (\alpha - 1) \frac{H_\mathrm{dry} k_B T}{\ell_K^3} \log\left( \frac{p_\mathrm{sat}}{p_\mathrm{tot} - p_\mathrm{sat}} \right).
\end{align}
It is apparent that for the observed swelling ratios and using $p_\mathrm{sat}\ll p_\mathrm{tot}$ the magnitude of the liquid bulk energy strongly supersedes the brush potential (yet, both are negative).

\bibliography{literature}

\newpage

\section*{Supplementary material}
\renewcommand\thefigure{S\,\arabic{figure}}
\renewcommand\theHfigure{S\,\arabic{figure}}
\setcounter{figure}{0}
\renewcommand{\theequation}{S\arabic{equation}}
\renewcommand{\theHequation}{S\arabic{equation}}
\setcounter{equation}{0}

\subsection{Interferometry analysis}

\begin{figure}[hbt!]
    \centering
    \includegraphics[width=\textwidth]{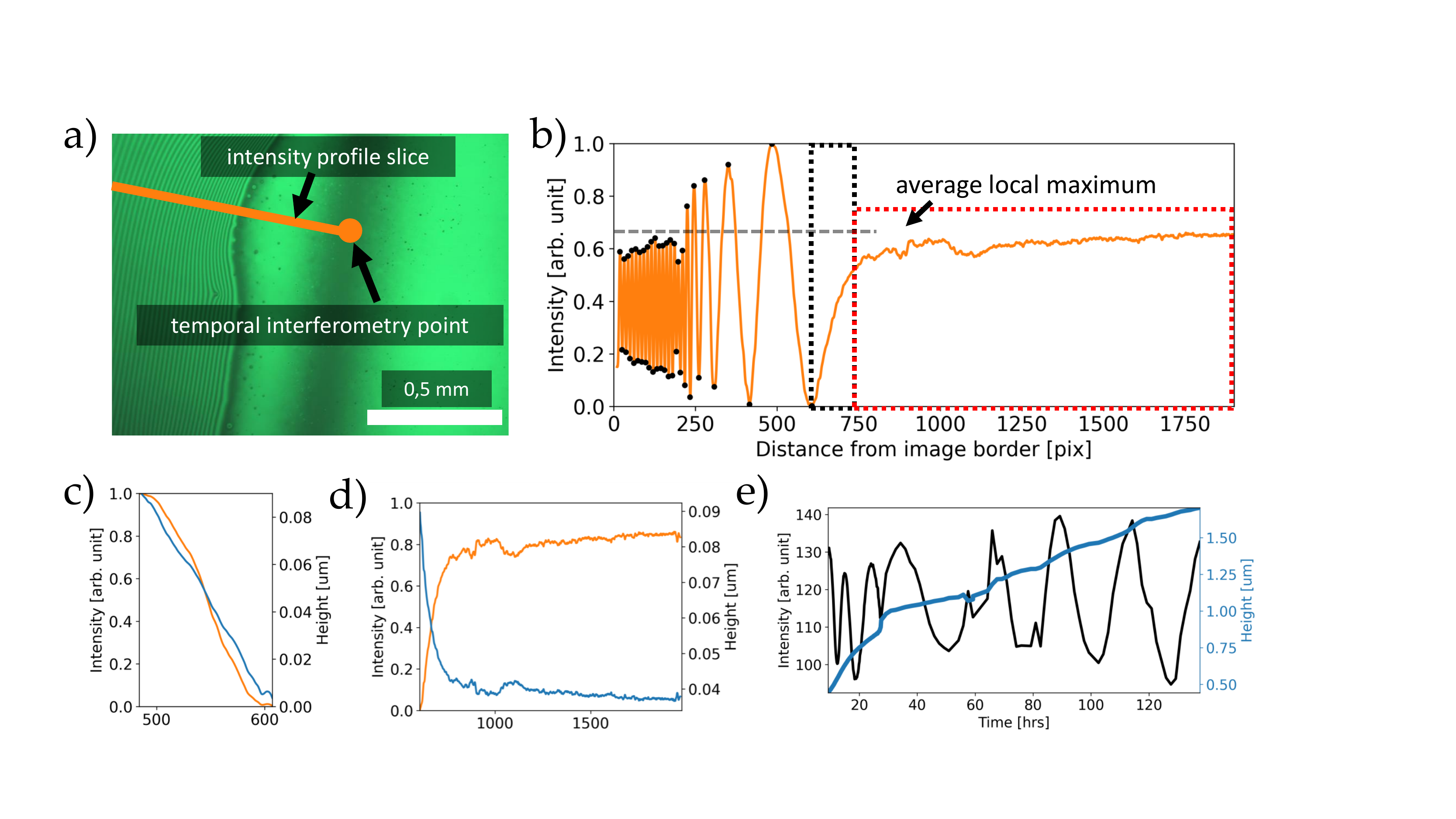}
    \caption{(a) An image of an interferometry pattern (HTK, closed cell, t$\approx$16 hours) as retrieved from the camera. The orange line indicates the location of the intensity profiles taken to calculate the average intensity profile. (b) Normalized intensity profile averaged from 15 parallel slices indicated in (a) (orange line), the local extrema (black dots) and the average local maximum (dash grey line). (c) Magnification of the black and (d) red bounded areas within (b) of averaged normalized intensity profile (orange) and the height profile (blue, right axis). (e) The intensity in time (black) of the fixed orange points is shown in (a) and the obtained height profile from this intensity profile (blue).}\label{fig:SI_1}
\end{figure}

Images of the interferometry pattern of the drop and halo are recorded every minute using a 24MP camera (Basler a2A5328 - 15ucBAS) ) fitted onto a microscope (Nikon Eclipse L150) fitted with a zoom objective (Nikon Plan UW 2x/0.06 $\infty$/- WD 7.5 OFN25, Nikon CFI Plan Fluor 4x/0.13 Phl DL 8/1.2 WD 16.4, Nikon LU Plan ELWD 20x/0.40 A $\infty$/0 V/D 13.0 EPI) and a 532nm bandpass filter (Thorlabs FL05532-10, \SI{532\pm 2}{nm}, FWHM = \SI{10\pm 2}{nm}). The recorded high-contrast images are analyzed using a homemade routine~\cite{githubwebsite}, written in Python 3.
First, the image (see Figure~\ref{fig:SI_1}(a)) is converted to a gray-scale image, and the contrast is enhanced using a Contrast Limited Adaptive Histogram Equalization algorithm from the NumPy library. Next, the intensity profiles of a set of parallel lines spaced 1 pixel apart perpendicular to the interferometry fringes are determined. Between 5 and 30 straight interferometry fringes are processed in this way. From the resulting slices, the normalized average intensity profile is calculated.

\subsubsection{Peak finding method}
For images recorded with a low magnification objective (4x) where the thickness of the halo is important, the profile is smoothened using a moving average filter and normalized. Using the built-in \texttt{findpeaks} algorithm, the local extrema (maxima and minima) are determined and corrected manually if necessary (see Figure~\ref{fig:SI_1}(b)).
The total height change $\Delta h$ from a reference point can be extracted from the interferometry pattern by counting the number of fringes:
\begin{equation}
    \Delta h = \frac{\lambda N}{2n},
\end{equation}
where $\lambda$ is the wavelength of the light (in vacuum), N is the total number of fringes, and n is the refractive index of the medium ($n_\mathrm{hexadecane}$=\num{1.4329}). Since the refractive indices of PLMA ($n_\mathrm{PLMA}$=\num{1.4740}) and hexadecane are relatively close, the change in the refractive index due to the liquid/brush composition changes during the spreading is disregarded, and $n_\mathrm{hexadecane}$ is used for the analysis. The intensity between the fringes changes with a cosine, thus relating the intensity between 2 local extrema directly with the height:
\begin{equation}
I = \frac{1}{2}\cos{\left[\frac{4\pi h}{\lambda/n}+1\right]}\Rightarrow h(x).
\end{equation}
The intensity profiles between all the extrema are fitted to this model, and the height profile as a function of lateral distance $x$ is calculated (see Figure~\ref{fig:SI_1}(c)). This approach does not work for the intensity profiles before the first local extrema (head) and after the last local extrema (tail). Ideally, all the local extrema are either 0 or 1, but due to non-ideal reflections and averaging errors, this is not the case. For the head and tail, it is thus unknown if a local extremum is already reached or what the intensity of the upcoming local extremum will be. Instead, it is assumed that the local extrema of the head and tail are the average of all other extrema in the intensity profile. These sections are then normalized with regard to this average local extremum, and the height profile is calculated using the same method mentioned above (see Figure~\ref{fig:SI_1}(d)).
The calculated height profiles between the local extrema are stitched together to form the total height profile of the intensity profile. 
The height profiles can be derived from the normalized average intensity profiles using either the peak finding method or the Fast Fourier Transform method.

\subsubsection{Fast Fourier Transform method}
Images recorded with a high magnification objective (20x), where more fringes within the drop are visible, and the contact angle of the drop is important, are transformed using the Fast Fourier Transform (FFT) to obtain the height profile.

First, the FFT of the intensity profile is calculated. Since most of the noise in the image is found in extremely high frequencies, these frequencies are removed in the Fourier domain of the intensity profile. 
Similarly, the low frequencies are removed as they mainly presents optical vignetting caused by the lens. Then, the filtered spectrum is transformed back using the inverse FFT, which results in a real and imaginary part. To obtain the phase of the fringe pattern, an \texttt{atan2}  operation is performed, where the \texttt{atan2(y,x)} is defined as the angle between the positive $x$-axis and the line from the origin to the point $(x,y)$. For a proper bandpass-filtered image, this results in a step-like function of period and amplitude of $2\pi $. Using a built-in unwrapping function from the NumPy library, the wrapped space is unwrapped, resulting in the height profile in units of $\pi$. The final height profile is then obtained by multiplying the unwrapped height profile by $\lambda \pi/n$.
Since the interferometric technique described above only provides relative changes in thickness, we determined the absolute thickness of the dry brush layers by ellipsometry. To trace the evolution of the absolute film thickness in zoomed interferometry images near the contact line, we followed the interferometry signal from a fixed point far away from the contact line (orange dot in Figure~\ref{fig:SI_1}(a)) in time and added the corresponding offsets to the position-dependent thickness profiles (see Figure~\ref{fig:SI_1}(e)).

\newpage
\subsubsection{Spreading over pre-swollen brushes}

\begin{figure}[hbt!]
    \centering
    \includegraphics[width=\textwidth]{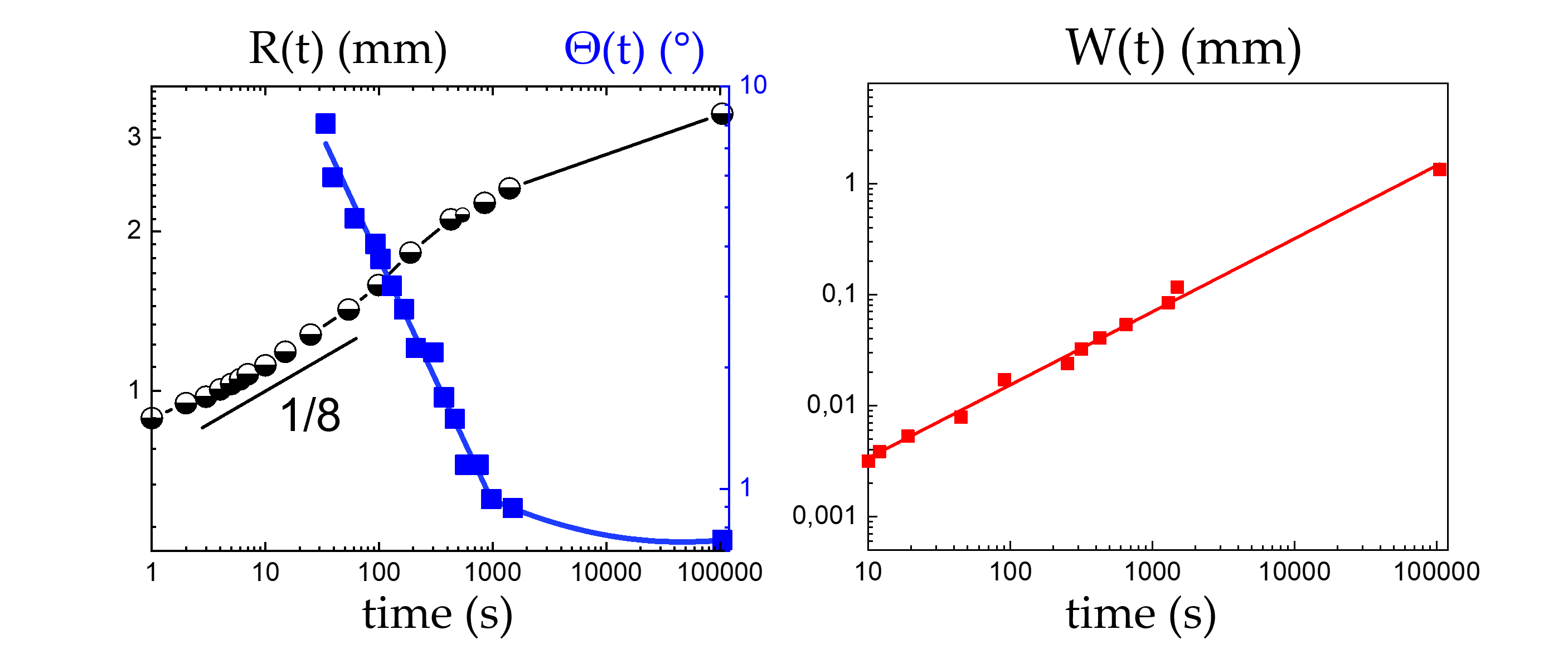}
    \caption{Characterization of macroscopic drop spreading on pre-saturated PLMA brush layer in the closed configuration saturated with HD vapor atmosphere. Left panel: drop radius  $R(t)$ (black) and contact angle $\theta(t)$ (blue). Right panel: halo width $W(t)$.}\label{fig:SI_2}
\end{figure}

\end{document}